\documentclass[sigconf]{acmart}
\renewcommand\footnotetextcopyrightpermission[1]{} 

\usepackage{fancyvrb}
\usepackage{multirow}
\usepackage{booktabs}
\usepackage{threeparttable}
\usepackage{ifthen}
\usepackage{color}
\usepackage{bbding}
\usepackage{makecell}
\usepackage{threeparttable}
\usepackage{amsmath}
\usepackage{stfloats} 
\usepackage{xspace}
\usepackage[T1]{fontenc}
\usepackage{paralist}
\usepackage{enumerate}
\usepackage[linesnumbered,ruled,vlined]{algorithm2e}
\usepackage{array, multirow,graphicx}
\usepackage{float}
\usepackage{balance}
\usepackage{tikz}
\usepackage{calc}
\usepackage{subfigure}
\usepackage{listings,amsfonts}
\usepackage{xcolor,pifont}
\usepackage{url}
\usepackage{hyperref, epsfig, endnotes}
\usepackage{makecell}
\usepackage[normalem]{ulem}
\usepackage[misc]{ifsym}
\usepackage{subcaption}
\usepackage{cleveref}
\usepackage{tcolorbox}
\usepackage{python}
\usepackage{graphicx}
\usepackage{float}
\usepackage{enumitem}

\definecolor{customblue}{HTML}{006ca6}
\definecolor{customgreen}{HTML}{009264}
\definecolor{custombrown}{HTML}{ff3d00}
\AtEndPreamble{
\usepackage{hyperref}

    \hypersetup{
      colorlinks = true,
      linkcolor = customblue,
      anchorcolor = purple,
      citecolor = customblue,
      filecolor = purple,
      urlcolor = customblue
    }
}

\newcommand{\find}[1]{
\begin{tcolorbox}[leftrule=0.5mm,toprule=0mm,bottomrule=0mm,left=0.7pt,right=0.7pt,top=0.2pt,bottom=0.2pt]
\em #1
\end{tcolorbox}
}

\lstdefinestyle{customstyle}{
    commentstyle=\color{green!50!black},
    keywordstyle=\color{purple!70!black},
    stringstyle=\color{orange!70!black},
    basicstyle=\ttfamily\footnotesize,
    breakatwhitespace=false,
    breaklines=true,
    captionpos=b,
    keepspaces=true,
    numbers=left,
    numbersep=5pt,
    showspaces=false,
    showstringspaces=false,
    showtabs=false,
    tabsize=2,
    frame=single,
    framesep=2pt,
    framerule=0.4pt,
    rulecolor=\color{gray!50},
}

\newcommand{\tool}{\textsc{DeclarUI}}

\begin{document}

\title{Bridging Design and Development with Automated Declarative UI Code Generation}

\author[T Zhou]{Ting Zhou}
\email{tingzhou27@hust.edu.cn}
\authornotemark[1]
\affiliation{%
  \institution{Huazhong University of Science and Technology}
  \city{Wuhan}           
  \country{China}
}
\author[Y Zhao]{Yanjie Zhao}
\email{Yanjie_Zhao@hust.edu.cn}
\authornote{Ting Zhou and Yanjie Zhao are the co-first authors.}
\affiliation{%
  \institution{Huazhong University of Science and Technology}
  \city{Wuhan}           
  \country{China}
}
\author[X Hou]{Xinyi Hou}
\email{xinyihou@hust.edu.cn}
\affiliation{%
  \institution{Huazhong University of Science and Technology}
  \city{Wuhan}           
  \country{China}
}
\author[X Sun]{Xiaoyu Sun}
\email{Xiaoyu.Sun1@anu.edu.au}
\affiliation{%
  \institution{The Australian National University}          
  \country{Australia}
}
\author[K Chen]{Kai Chen}
\email{kchen@hust.edu.cn}
\authornotemark[2]
\affiliation{%
  \institution{Huazhong University of Science and Technology}
  \city{Wuhan}           
  \country{China}
}
\author[H Wang]{Haoyu Wang}
\email{haoyuwang@hust.edu.cn}
\authornote{Corresponding authors.}
\affiliation{%
  \institution{Huazhong University of Science and Technology}
  \city{Wuhan}           
  \country{China}
}

\begin{abstract}
Declarative UI frameworks have gained widespread adoption in mobile app development, offering benefits such as improved code readability and easier maintenance. Despite these advantages, the process of translating UI designs into functional code remains challenging and time-consuming. Recent advancements in multimodal large language models (MLLMs) have shown promise in directly generating mobile app code from user interface (UI) designs. However, the direct application of MLLMs to this task is limited by challenges in accurately recognizing UI components and comprehensively capturing interaction logic. 

To address these challenges, we propose \tool{}, an automated approach that synergizes computer vision (CV), MLLMs, and iterative compiler-driven optimization to generate and refine declarative UI code from designs. \tool{} enhances visual fidelity, functional completeness, and code quality through precise component segmentation, Page Transition Graphs (PTGs) for modeling complex inter-page relationships, and iterative optimization. In our evaluation, \tool{} outperforms baselines on React Native, a widely adopted declarative UI framework, achieving a 96.8\% PTG coverage rate and a 98\% compilation success rate. Notably, \tool{} demonstrates significant improvements over state-of-the-art MLLMs, with a 123\% increase in PTG coverage rate, up to 55\% enhancement in visual similarity scores, and a 29\% boost in compilation success rate. We further demonstrate \tool{}'s generalizability through successful applications to Flutter and ArkUI frameworks. User studies with professional developers confirm that \tool{}'s generated code meets industrial-grade standards in code availability, modification time, readability, and maintainability. By streamlining app development, improving efficiency, and fostering designer-developer collaboration, \tool{} offers a practical solution to the persistent challenges in mobile UI development.

\end{abstract}

\maketitle

\section{Introduction}

As the mobile app ecosystem continues to expand, with a constant influx of new apps entering the market, the development of user interfaces (UIs) has evolved to meet the increasing demands for better user experiences and more efficient development processes. Traditional imperative UIs, which specify step-by-step instructions for rendering UI elements, have given way to declarative UIs, which describe the desired state of the UI rather than the sequence of operations to achieve it~\cite{wolfpack2024,increment2024}. Declarative UIs offer several advantages over their imperative counterparts, such as improved maintainability, testability, and separation of concerns by decoupling the UI logic from the underlying implementation details.

The adoption of declarative UI frameworks in mobile app development has become widespread due to these benefits~\cite{increment2024,flatirons2024}. However, while these frameworks have simplified many aspects of UI development, they have not significantly reduced the amount of manual coding required. In real-world scenarios, the development process typically begins with the design of the UI~\cite{mobileappdaily2024}, followed by the complex and time-consuming task of translating these visual designs into functional declarative UI code. This process requires developers to manually map visual elements to their corresponding implementation details, which remains prone to errors and can be particularly challenging despite the advantages of declarative frameworks. Consequently, \textbf{there is a need for automated tools to streamline declarative UI code generation directly from UI designs}.

Despite this evident need, current research efforts in this area are still inadequate. Some methods rely on heuristics or machine learning models to extract UI components and layout information from design images~\cite{object_detect,GUI_skeleton}, but they often struggle with complex designs and lack ability to handle interactive logic. Other approaches use program synthesis techniques to generate code from natural language descriptions~\cite{wu2024uicoderfinetuninglargelanguage}, but they require detailed specifications and cannot directly interpret visual designs. Researchers have also explored generating UI code across different frameworks~\cite{beltramelli2017pix2codegeneratingcodegraphical,encoder-decoder}, but these approaches are carried out on traditional imperative UI frameworks and cannot be extended to declarative UI frameworks. In summary, \textbf{existing research falls short in addressing specific challenges of automatically generating declarative UI code for real-world development scenarios and lacks ability to generalize across different UI frameworks}.

Recent advancements in multimodal large language models (MLLMs) have opened up new possibilities for generating code from visual inputs. MLLMs, which integrate image processing capabilities into large language models (LLMs), have demonstrated superior performance in understanding and interpreting visual information compared to traditional computer vision (CV) models based on convolutional neural networks (CNNs)~\cite{yang2023dawnlmmspreliminaryexplorations,wu2023visualchatgpttalkingdrawing,chang2023statisticalinferencescomplexdependence}. Furthermore, LLMs have exhibited remarkable proficiency in various code intelligence tasks, such as code generation~\cite{codegeneratesurveyevaluatinglargelanguage,HumanEval,jiang2024surveylargelanguagemodels,lu2021codexgluemachinelearningbenchmark}, translation~\cite{kc2023neuralmachinetranslationcode,deng2024raconteurknowledgeableinsightfulportable}, summarization~\cite{arakelyan2023exploringdistributionalshiftslarge,chen2022transferabilitypretrainedlanguagemodels,Gao2023ConstructingEI,gu2022assemblefoundationmodelsautomatic}, and repair~\cite{liu2024marscodeagentainativeautomated,dehghan2024mergerepairexploratorystudymerging,xu2024automatedccprogramrepair,yang2024crefllmbasedconversationalsoftware}. The synergy between visual comprehension and code generation abilities within MLLMs presents a promising avenue for translating UI designs into functional code, bridging the gap between visual representations and programmatic implementations.

Despite the powerful image understanding and code generation abilities of MLLMs, directly applying them to UI code generation faces several significant challenges. \textbf{Imprecise component recognition} often occurs due to the complexity of UI designs and the lack of explicit annotations. MLLMs also struggle with \textbf{a limited understanding of interactive logic}, failing to accurately infer navigation flow and event handling from static images. Furthermore, they often generate \textbf{inconsistent behavior across pages}, producing code that may work for individual screens but fails to maintain coherence in a multi-page app. Lastly, the generated code frequently suffers from \textbf{reliability and compilation issues}, containing syntax errors or violating framework-specific conventions. These challenges underscore the need for a specialized approach that can bridge the gap between MLLMs' capabilities and the specific requirements of declarative UI code generation, addressing accuracy, consistency, and reliability in the process.

To fill this gap, we propose \tool{}, a novel approach to enhance MLLM-based declarative UI code generation from UI designs. \tool{} integrates advanced CV techniques with MLLMs to decompose complex UI designs into structured component information. We introduce the Page Transition Graph (PTG) to represent the app's navigation logic, which, combined with the structured component information, forms a comprehensive prompt for the MLLM. After initial code generation, \tool{} employs an iterative refinement process, performing navigational integrity and compilation checks to identify and rectify common errors. This multi-stage approach ensures the generated UI code is both visually accurate and functionally robust across multiple pages.

In summary, our key contributions are as follows:
\begin{itemize}
    \item We propose \tool{}, an approach that enhances MLLM-based declarative UI code generation from UI designs. \tool{} combines CV techniques with MLLMs to decompose UI designs into structured component information, introduces the PTG to represent navigation logic, and employs an iterative refinement process.

    \item Our study compiled a dataset of 50 top-ranked UI design sets (50\% mockups, 50\% app screenshots) from design websites and app stores across 10 categories. For React Native~\cite{reactnative2024}, \tool{} significantly outperformed state-of-the-art MLLMs like Claude-3.5~\cite{anthropic_claude_3_5}\footnote{The specific reference to ``Claude-3.5'' in this paper refers to \textbf{Claude 3.5 Sonnet}.} and GPT-4o~\cite{openai_hello_gpt4o} in generating UI code. \tool{} achieved a 123\% improvement in PTG coverage rate, up to 55\% increase in visual similarity scores, and a 29\% higher compilation success rate. Furthermore, \tool{} demonstrated strong generalization to Flutter~\cite{flutter2024} and ArkUI~\cite{huawei2024}, effectively bridging UI designs and functional declarative UI code across frameworks. 

    \item We conducted a user study with professional developers to evaluate the performance of \tool{}. The results demonstrate that \tool{} significantly outperforms the baseline in terms of code availability (4.97 vs. 4.15), modification time (4.03 vs. 2.45), readability (4.62 vs. 3.75),  and maintainability (4.55 vs. 3.37). Developers reported that the UI code generated by \tool{} meets industrial-grade standards in multiple dimensions, highlighting its potential to streamline app development processes and improve designer-developer collaboration.

\end{itemize}

\section{Background and Motivations}

\begin{figure}[ht!]
  \centering
  \subfigure[Screenshots of an app named ``com.lazada.android''.]{
  \includegraphics[width=\linewidth]{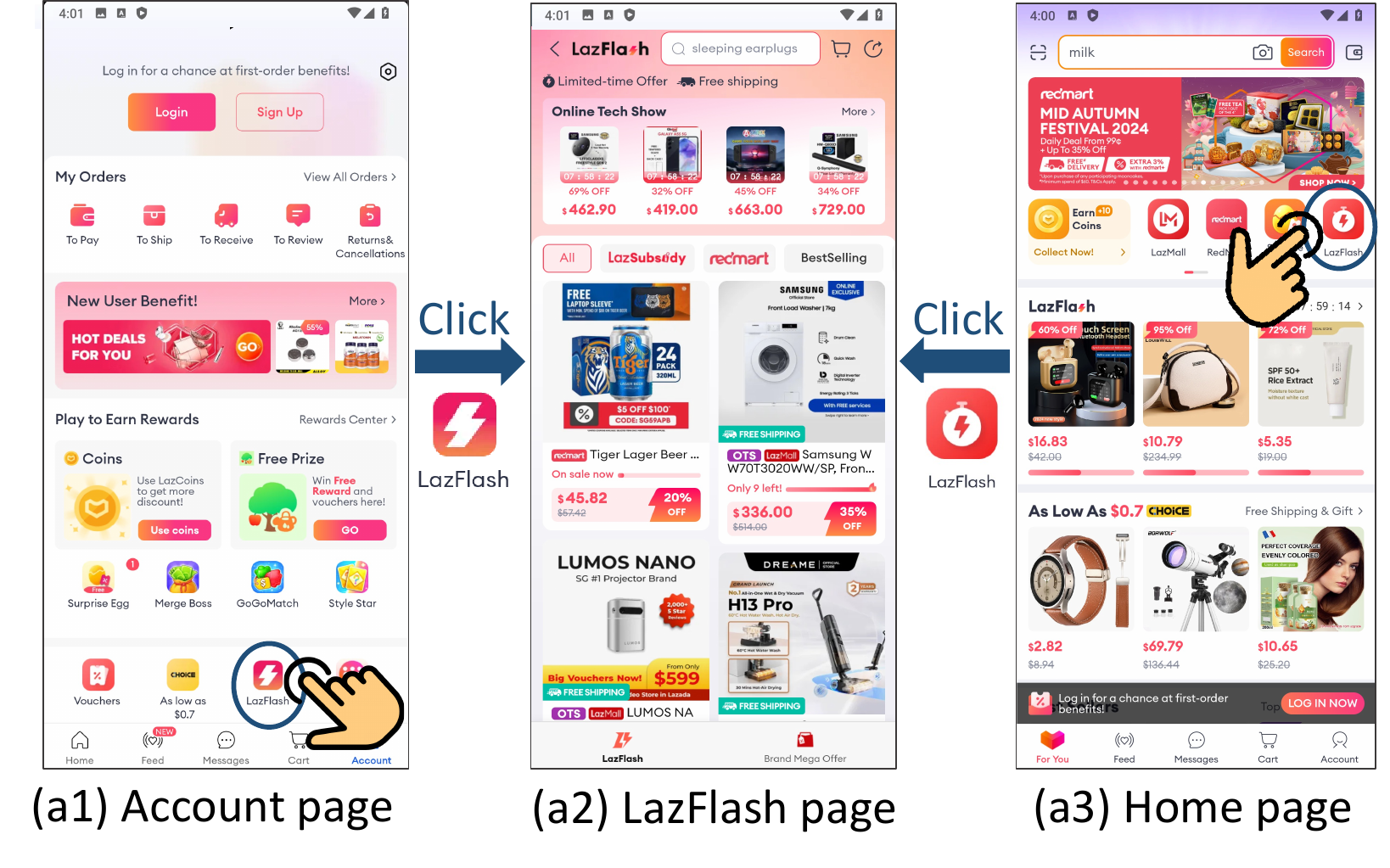}
  \label{fig:motivating original}
  }
  \hspace{0.1cm}
  \subfigure[UI rendering of the generated UI code by Claude-3.5.]{
  \includegraphics[width=\linewidth]{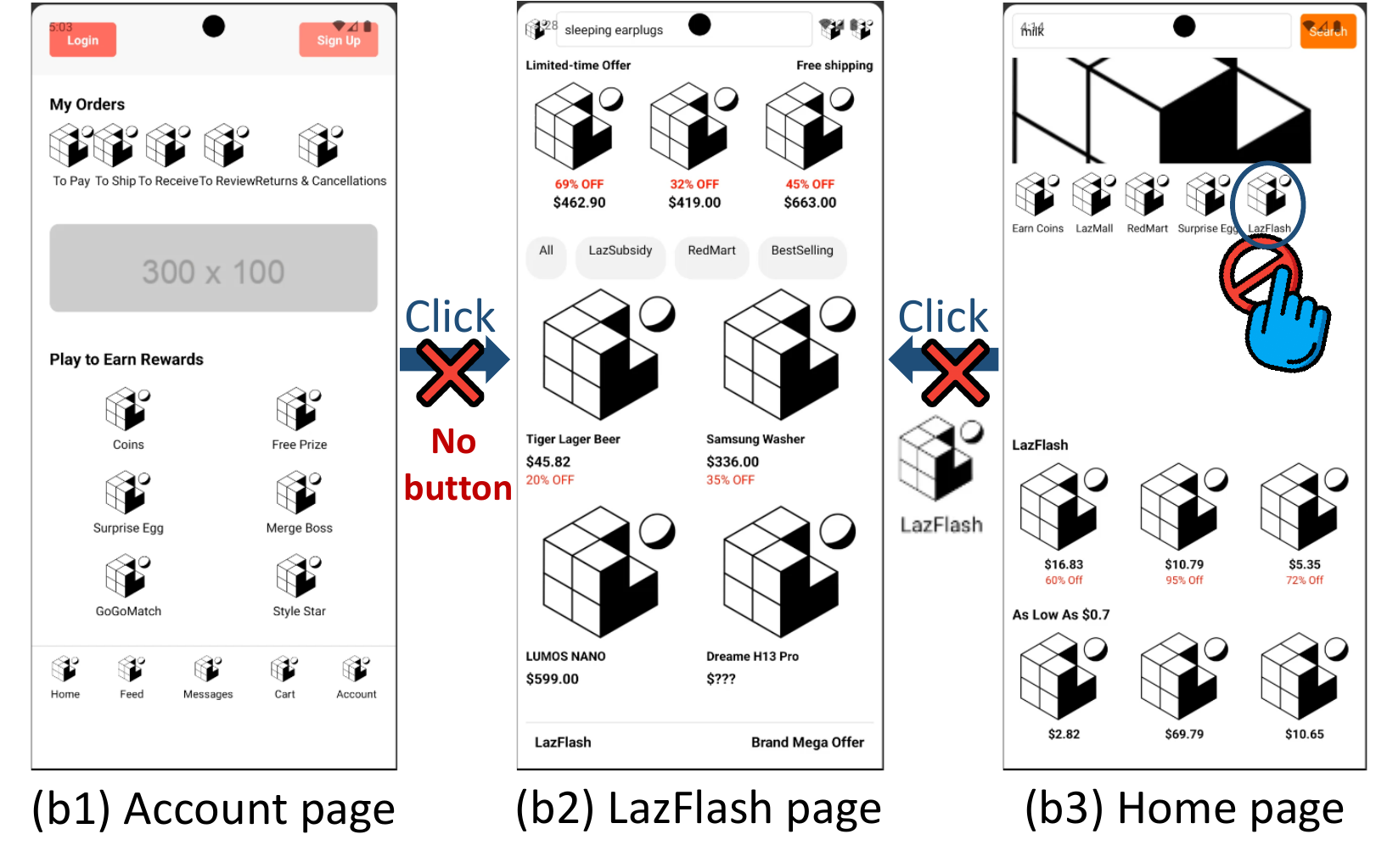}
  \label{fig:motivating raw llm}
  }
  \caption{Motivating example comparing the original app screenshots with the rendering of code generated by an MLLM. Please note that this paper focuses on code generation evaluation and does not consider the impact of image assets on the generation process.}
  \label{fig:motivating}
\end{figure}

The rapid advancement of MLLMs has ushered in a new era of possibilities for automating mobile app development. Models such as OpenAI's DALL-E~\cite{ramesh2021zeroshottexttoimagegeneration} and Google's Imagen~\cite{saharia2022photorealistictexttoimagediffusionmodels} have showcased remarkable abilities in generating images and text from natural language prompts, inspiring researchers to explore their potential in generating app code directly from UI design mockups or app screenshots~\cite{gui2024vision2uirealworlddatasetlayout,wan2024automaticallygeneratinguicode,xiao2024prototype2codeendtoendfrontendcode}. This innovative approach holds the promise of streamlining the app development process, potentially allowing designers to transform their visual concepts into functional UI code with unprecedented ease and speed.

Unfortunately, despite the exciting potential of MLLMs for UI code generation, current approaches face significant challenges when processing raw UI design mockups or screenshots. In \autoref{fig:motivating}, we provide an example of a shopping app named ``com.lazada. android''~\cite{LazadaApp2024}, sourced from Google Play, to illustrate some of the issues with generating UI code using an MLLM, i.e., Claude-3.5:

\noindent\underline{\textbf{Imprecise Component Recognition.}}
The MLLM struggles with accurate identification and rendering of UI components. As evident in \autoref{fig:motivating}(b1), the generated \textit{Account page} lacks several crucial elements present in the original design (\autoref{fig:motivating}(a1)). Most notably, the \textit{LazFlash button} is missing, which is a critical component for navigation within the app. This omission demonstrates that MLLMs may not fully capture all the necessary visual elements, potentially leading to incomplete or inaccurate UI implementations.

\noindent\underline{\textbf{Limited Understanding of Interactive Logic.}}
UI design encompasses more than just visual elements; it includes complex interaction patterns and user experience considerations. The MLLM faces difficulties in accurately interpreting the designer's intentions and the relationships between different elements. This is clearly illustrated by the lack of navigation logic in the generated UI code. The arrows with crosses in \autoref{fig:motivating raw llm} indicate that the expected transitions between pages (such as from the \textit{Account page} to the \textit{LazFlash page}) are not implemented in the generated code, despite being clearly present in the original designs shown in \autoref{fig:motivating original}.

\noindent\underline{\textbf{Inconsistent Behavior Across Pages.}}
The MLLM struggles to recognize and implement consistent behaviors across different pages of an app. This challenge is exemplified in \autoref{fig:motivating}, where both the \textit{Account page }(a1) and the \textit{Home page} (a3) should navigate to the \textit{LazFlash page} (a2) when the \textit{LazFlash button} is clicked. However, the generated UI code fails to maintain this consistency, as indicated by the crossed-out arrows in \autoref{fig:motivating raw llm}. This suggests that MLLMs have difficulty in analyzing and implementing complex attention patterns across multiple screenshots, potentially leading to inconsistent user experiences in the generated app.

\noindent\underline{\textbf{Code Reliability and Compilation Issues.}}
The output from the MLLMs can be unpredictable, often resulting in code that contains compilation errors or runtime issues. While not directly visible in \autoref{fig:motivating}, this challenge is implicit in the discrepancies between the original app screenshots and the rendered UI from the generated code. These issues necessitate significant human intervention to correct errors and ensure the code is functional, potentially offsetting the time-saving benefits of using MLLMs for UI code generation.

These challenges underscore the current limitations of MLLM-based approaches in UI code generation. While the potential for automating app development is promising, significant improvements in component recognition, understanding of interactive logic, cross-page consistency, and code reliability are necessary before MLLMs can reliably transform visual designs into fully functional UI code. Addressing these challenges will be crucial for realizing the vision of streamlined app development through AI-assisted code generation.

\textbf{Our Solutions.}
To address the issue of imprecise component recognition, we leverage \textbf{advanced CV techniques}. By integrating the Grounding DINO object detection model~\cite{liu2023grounding} with the Segment Anything image segmentation model~\cite{kirillov2023segany}, we are able to precisely isolate individual components within UI designs. This preprocessing step significantly enhances the MLLM's ability to recognize UI elements, ensuring that the generated code more accurately reflects the design intent.
To tackle the limitations in understanding interactive logic and maintaining consistent behavior across pages, we introduce the concept of \textbf{Page Transition Graph (PTG)}. The PTG serves as a structured representation that clearly delineates the navigation logic between different pages. By providing this high-level abstraction to the MLLM, we compensate for the model's weaknesses in comprehending complex inter-page relationships, thereby generating interaction logic that more closely aligns with the designer's intentions.
Finally, to improve code reliability and reduce compilation errors, we implement \textbf{an iterative refinement process}. This mechanism not only checks the navigational integrity of the MLLM-generated code but also performs compilation checks, ensuring consistent code quality. This approach significantly reduces the need for manual intervention, lowering time costs while simultaneously improving the overall quality of the generated code.

\section{Approach}

\begin{figure*}[h]
\centering
\includegraphics[width=0.9\textwidth]{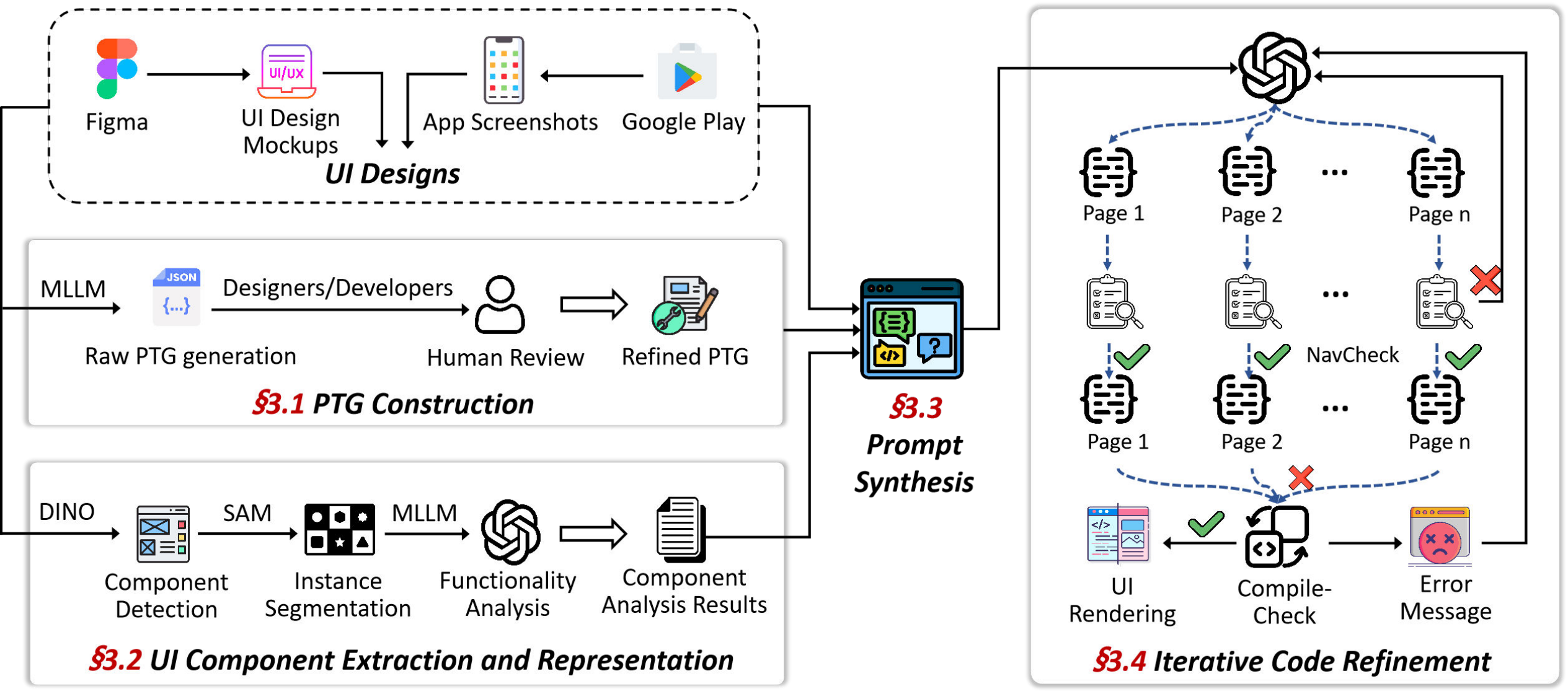}
\caption{The workflow of \tool{}.}
\label{fig:overview}
\end{figure*}

The workflow of \tool{}, as illustrated in \autoref{fig:overview}, starts with UI designs as input. These designs include high-quality Figma~\cite{figma2024} design mockups and screenshots of Google Play~\cite{GooglePlayStore2024} apps. 
Design mockups typically include both visual elements and text-based specifications, providing comprehensive design information. While text-based design specifications are readily interpretable by LLMs, \tool{} aims to address more challenging scenarios where \textbf{only visual designs are available}. Therefore, \tool{} processes both sources uniformly as image data, focusing exclusively on visual content. This unified image-based analysis allows \tool{} to process diverse inputs consistently, whether they are design mockups or app screenshots.

\tool{} preprocesses the input through two key steps: \textbf{PTG Construction} (\autoref{section:ptg construct}), which captures navigation logic, and \textbf{UI Component Extraction and Representation} (\autoref{section:component extract}), which utilizes CV techniques. Subsequently, the \textbf{Prompt Synthesis} step (\autoref{section:prompt synthesis}) integrates the preprocessed data—including the generated PTG, component analysis results, and the complete UI screenshot—into a comprehensive prompt. This prompt serves as the foundation for the MLLM to generate complete UI code. Lastly, an \textbf{Iterative Code Refinement} process (\autoref{section:refinement}) checks for errors and navigation absence, ensuring the reliability of the final code output.

\subsection{PTG Construction}\label{section:ptg construct}

To address challenges rooted in the complexities of modern UI designs and MLLM limitations, we propose the utilization of PTGs as a key tool for capturing and representing UI interaction logic. We formally define a PTG where nodes represent individual app pages, and edges represent transitions between pages along with their triggering conditions. The \emph{PTG} is defined as a tuple:

\begin{equation*}
    PTG = (N, E)
\end{equation*}

where:

\begin{itemize}
    \item $N$ is a finite set of nodes representing app pages. Each node includes an $id$ as a unique identifier, a $name$ for the page name, a $type$ indicating the page type, and $property$ containing additional properties. $N$ is defined as:
    \begin{equation*}
        N = \{n_i = (id_i, name_i, type_i, property_i) \mid i = 1, 2, \ldots, k\}
    \end{equation*}
    
    \item $E$ is a set of directed edges representing transitions between pages. Each edge has specific attributes: $id$ is a unique identifier for the edge, \( ns, nt \in N \) denote the source and target nodes, while $condition$ is the condition that triggers the transition, and $action$ is the operation performed during the transition. $E$ is defined as:
    \begin{equation*}
        E = \{e_i = (id_i, ns_i, nt_i, condition_i, action_i) \mid i = 1, 2, \ldots, m\}
    \end{equation*}
    
\end{itemize}

\begin{figure*}[ht!]
\centering
\includegraphics[width=0.8\linewidth]{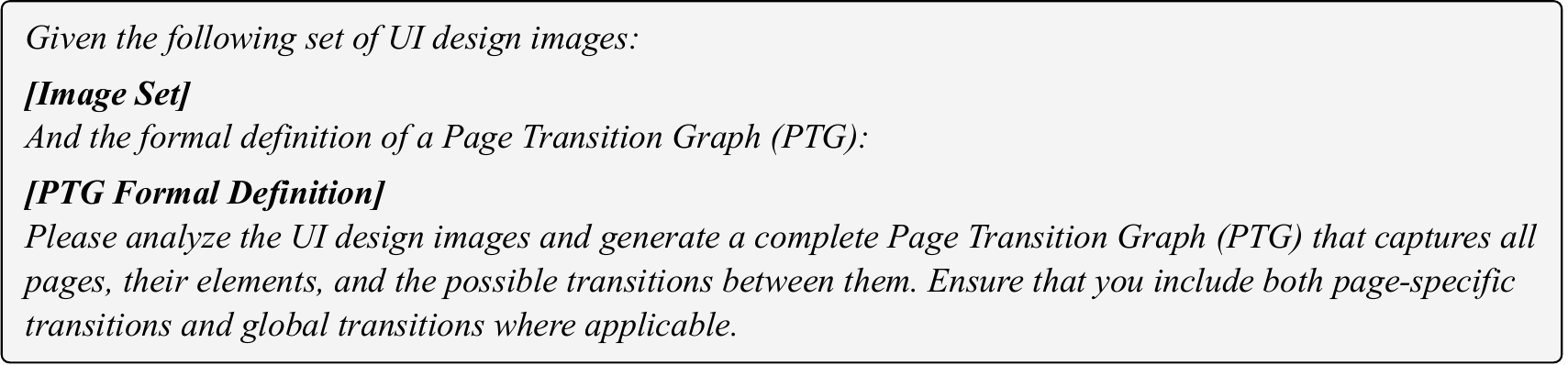}
\caption{Example of prompt used to construct PTG.}
\label{fig:ptg construct prompt}
\end{figure*}

\begin{figure}[ht!]
    \centering
    \includegraphics[width=\linewidth]{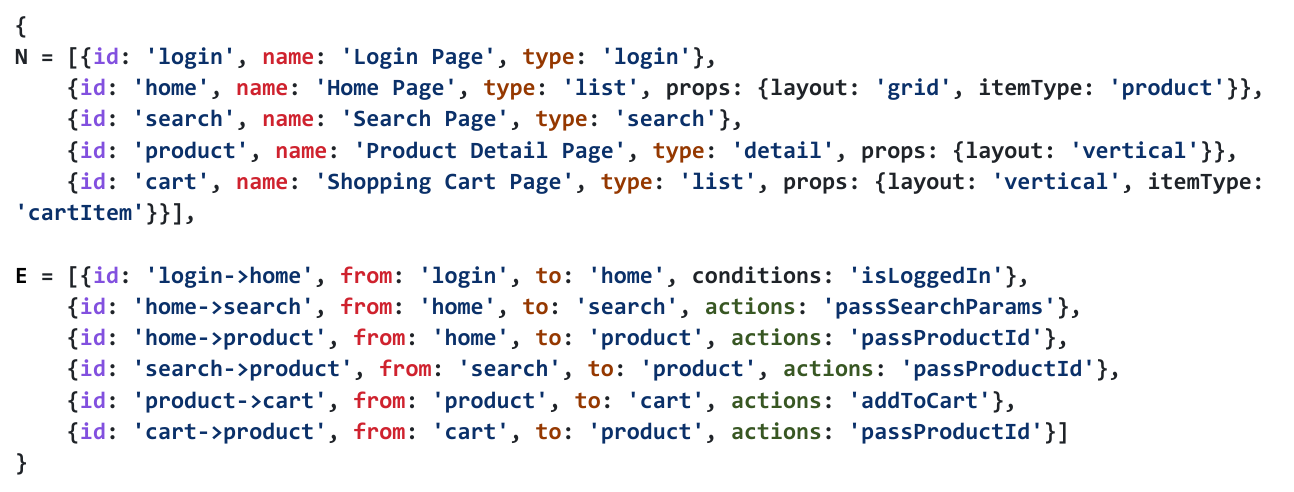}
    \caption{Example of the PTG for a simplified e-commerce app scenario.}
    \label{fig:ptg_example}
\end{figure}

Our introduction of PTGs seeks to offer MLLMs a clear, structured representation of interaction logic, effectively bridging the gap in expressing dynamic interactions in static UI designs. The PTG generation process leverages the robust inference capabilities of MLLMs combined with the analysis of UI designs. Initially, a set of UI designs is provided to an MLLM, i.e., Claude-3.5, accompanied by a formal definition of the PTG structure. The MLLM analyzes the UI designs and generates a comprehensive PTG based on the provided prompt, as exemplified in \autoref{fig:ptg construct prompt}. It is guided to output its analysis in a structured JSON format to facilitate further processing and integration. \autoref{fig:ptg_example} illustrates an example PTG for a simplified e-commerce app scenario, designed to clearly demonstrate the PTG concept. Finally, designers or developers review the generated PTG, making necessary corrections or additions to ensure the accuracy and completeness of the captured interaction logic.

PTGs offer a structured representation of UI elements, encompassing page hierarchies, transition logic, and navigation rules. PTGs play a crucial role in addressing key challenges associated with utilizing MLLMs for UI code generation from designs, including limited understanding of interactive logic and inconsistent behavior across pages.

\subsection{UI Component Extraction and Representation}\label{section:component extract}
\begin{figure}[ht!]
\centering
\subfigure[Original UI screenshot]{
\label{fig:original}
\includegraphics[width=0.2\linewidth]{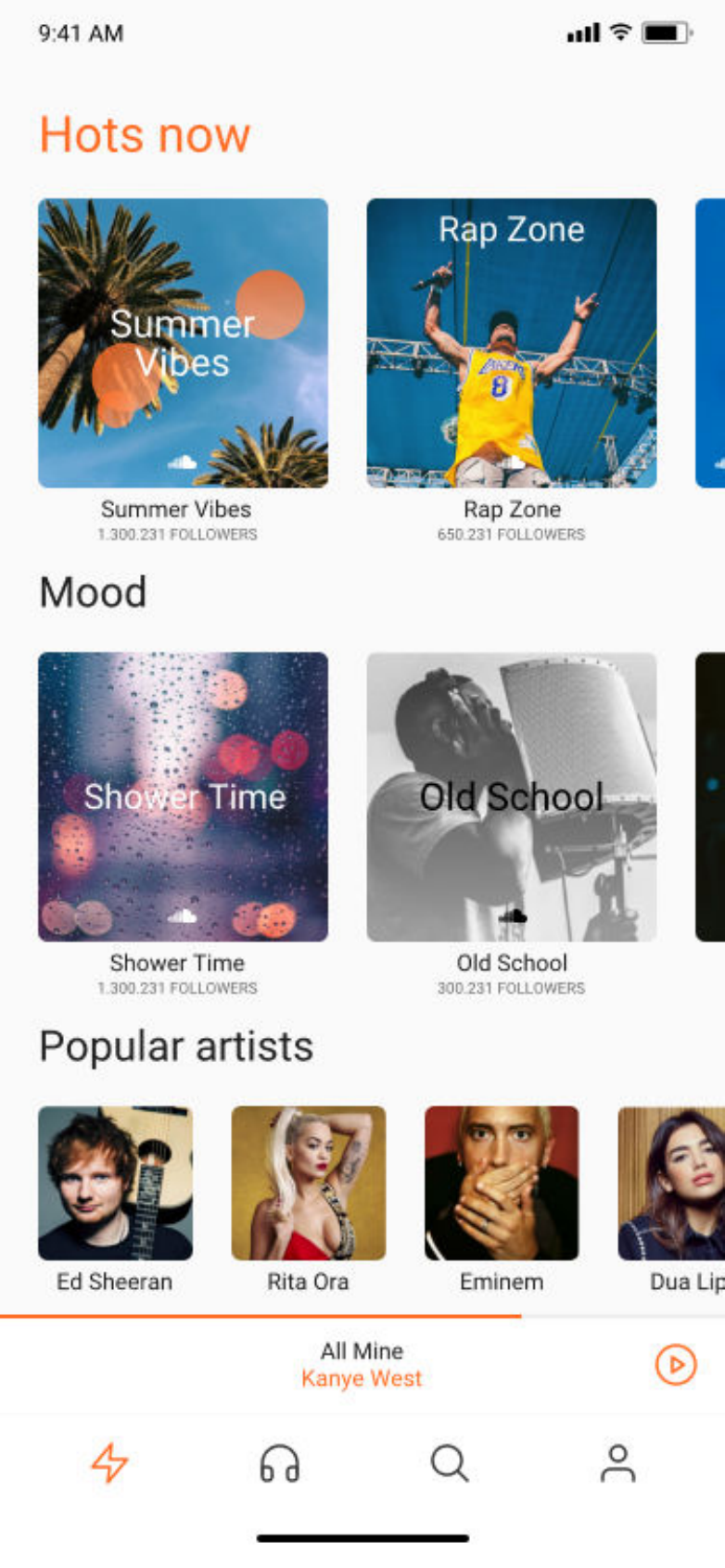}}
\hspace{0.1cm}
\subfigure[DINO-dectected UI]{
\label{fig:dino_ui}
\includegraphics[width=0.2\linewidth]{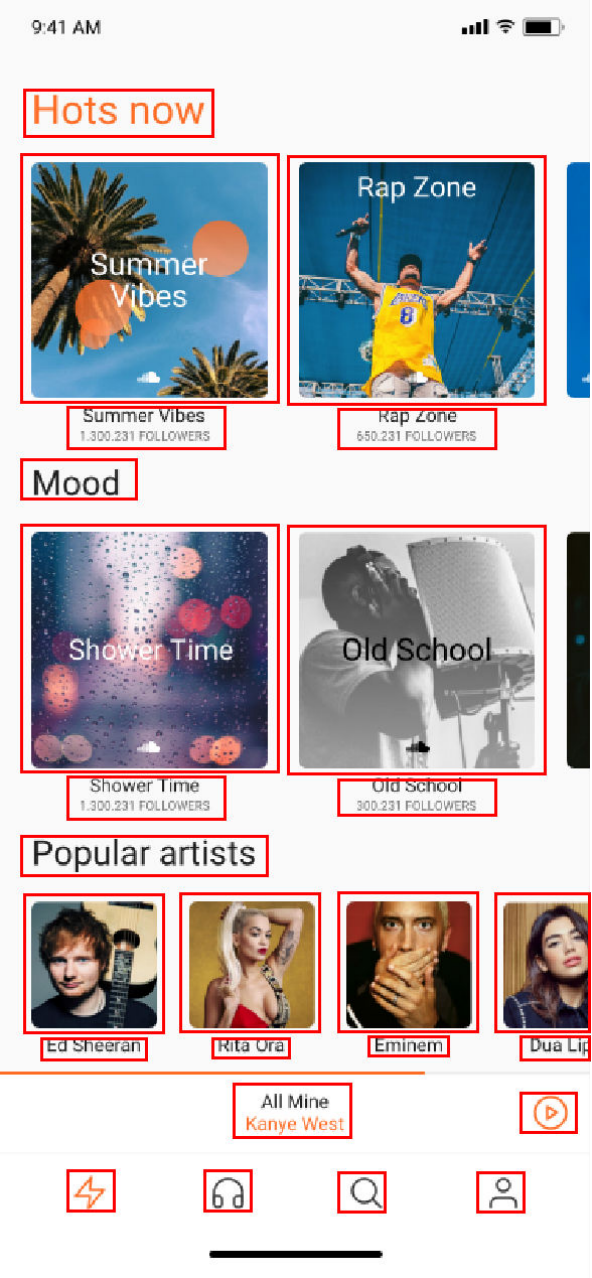}}
\hspace{0.1cm}
\subfigure[SAM-segmented UI]{
\label{fig:sam_ui}
\includegraphics[width=0.2\linewidth]{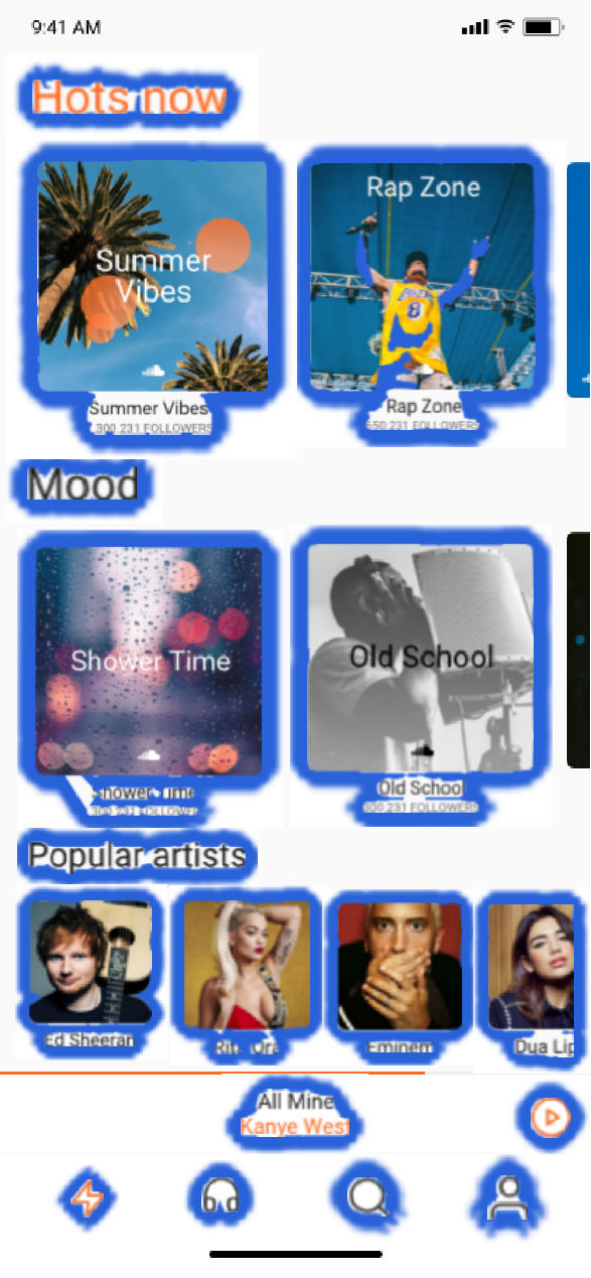}}
\hspace{0.1cm}
\subfigure[Component classification and functional analysis]{
\label{fig:component_classification}
\includegraphics[width=0.22\linewidth]{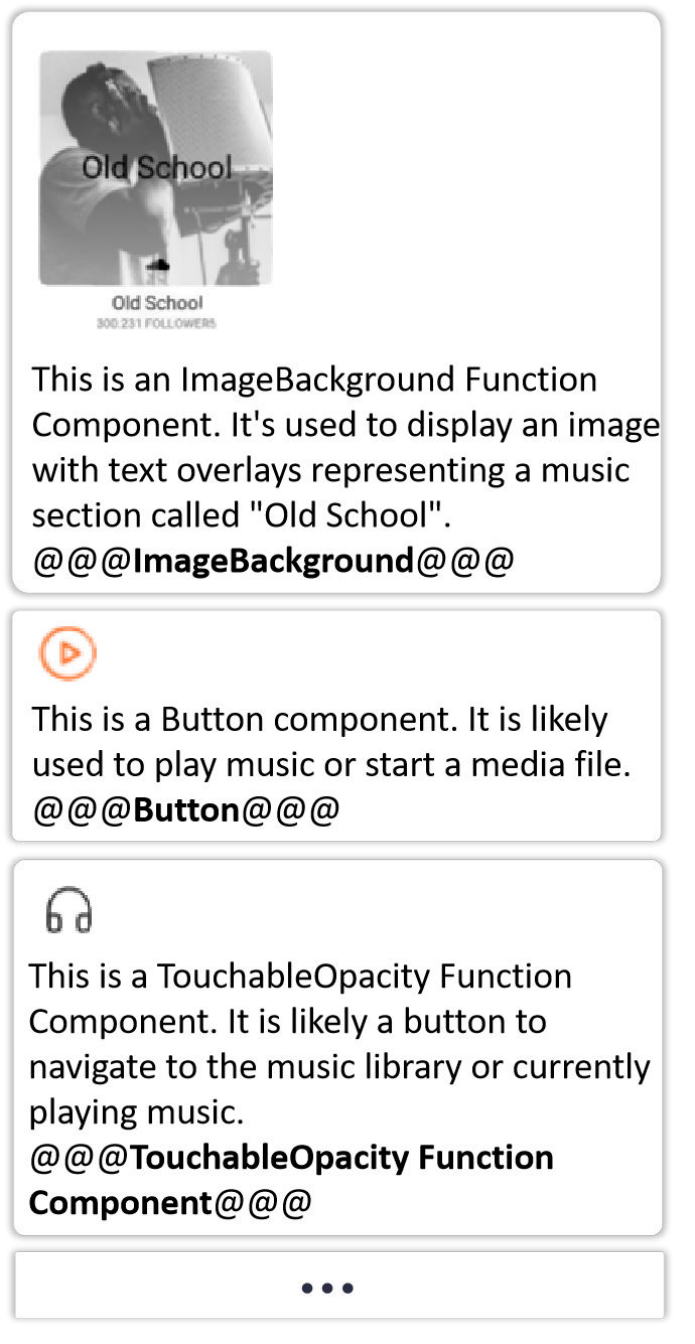}}
\caption{Example of the UI component extraction process.}
\label{fig:component_extact}
\end{figure}

This process aims to transform the complex UI design images into a structured representation of components, which is more easily understood and processed by MLLMs for generating high-quality code, as illustrated in \autoref{fig:component_extact}. It includes three stages as follows:

\noindent\textbf{Automated Component Detection.}
Instead of relying on manual annotation, \tool{} leverages the capabilities of Grounding DINO~\cite{liu2023grounding}, a state-of-the-art object detection model. Grounding DINO excels in zero-shot object detection, making it ideal for identifying potential UI components without requiring extensive training on UI-specific datasets. The model generates bounding boxes around detected objects, which serve as initial component localization.

\noindent\textbf{Automated Instance Segmentation.}
To achieve precise instance segmentation of the UI components, \tool{} employs the Segment Anything Model (SAM)~\cite{kirillov2023segany}. The bounding boxes produced by Grounding DINO act as prompts for SAM, guiding the segmentation process. This approach combines the strengths of both models: Grounding DINO's ability to detect objects without prior training, and SAM's capability to produce highly accurate segmentation masks.

\begin{figure}[ht!]
  \centering
  \subfigure[Example of the prompt used to extract component types.]{
  \includegraphics[width=2.8in]{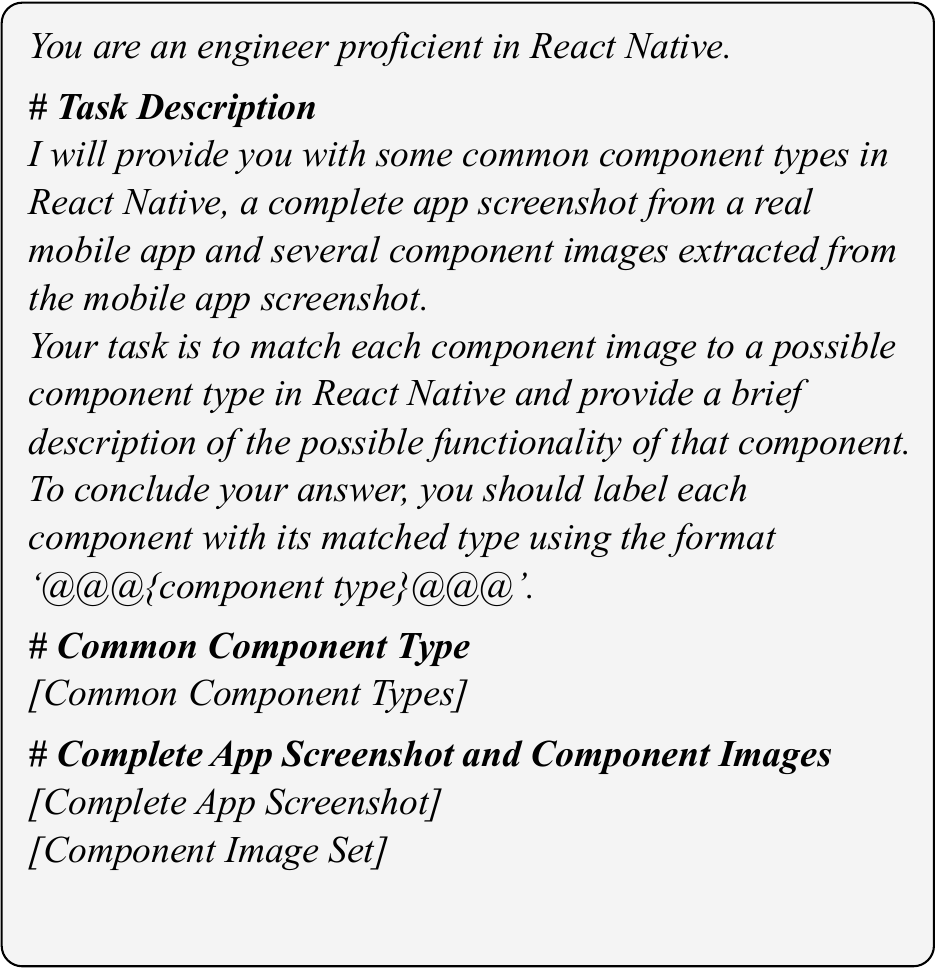}
  \label{fig:component_type_prompt}
  }
  \subfigure[Example of a synthesized prompt.]{
  \includegraphics[width=2.8in]{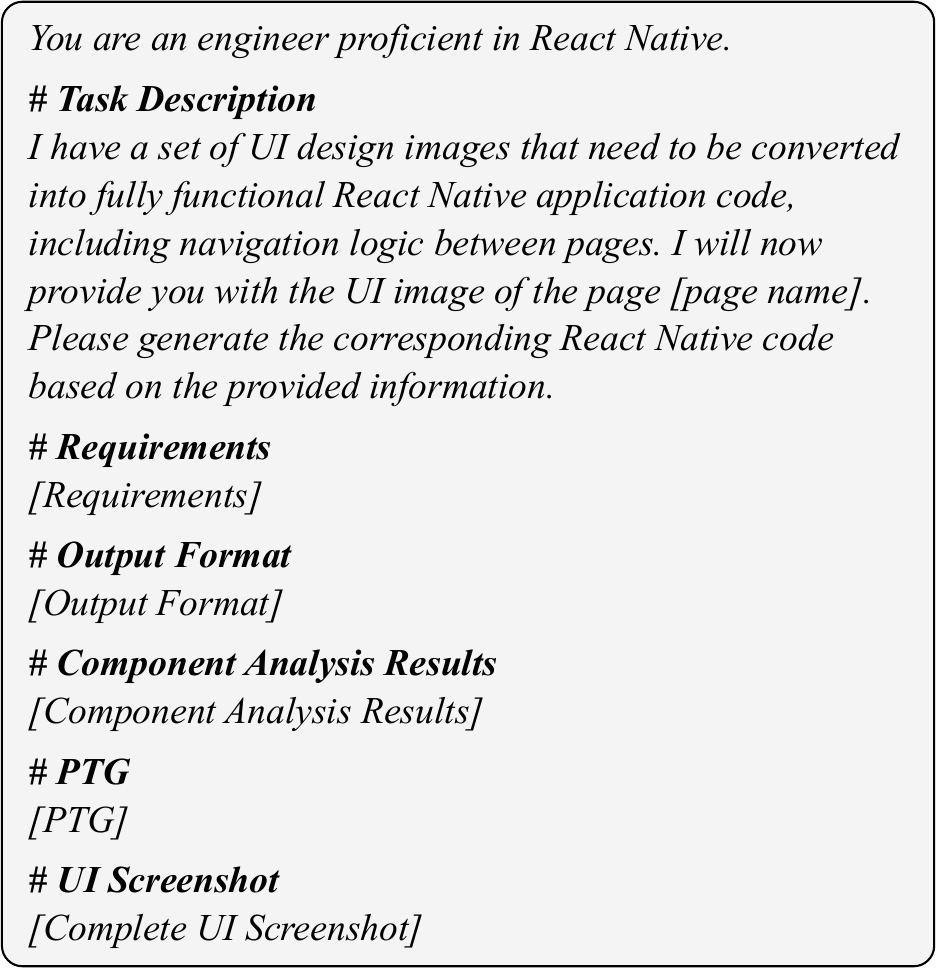}
  \label{fig:synthesized_prompt}
  }
  \caption{Examples of the prompts used in \autoref{section:component extract} and \autoref{section:prompt synthesis}.}
  \label{fig:prompt}
\end{figure}

\noindent\textbf{Component Classification and Functional Analysis.}
We then leverage an MLLM (i.e., Claude-3.5) to process the segmented component images. As shown in \autoref{fig:component_type_prompt}, a carefully designed prompt is used to extract the component type. The prompt includes the segmented images and instructs the MLLM to classify the component type (e.g., button, text field, dropdown menu, etc.) and infer its potential functionality within the UI. This step utilizes the MLLM's understanding of UI design patterns and its ability to reason about the purpose of visual elements.

The output of this three-stage process is a comprehensive, structured representation of the UI design images. This representation includes precise spatial information and instance segmentation masks for each component, type classifications, and inferred functional descriptions.

\subsection{Prompt Synthesis}
\label{section:prompt synthesis}

Prompt synthesis integrates the preprocessed data into a comprehensive prompt for the MLLM, enhancing its ability to generate UI code that accurately reflects both the visual elements and the underlying interactive logic of the original design. This process combines three key elements: the \textbf{PTG}, which provides a structured representation of the navigation logic and inter-page relationships; \textbf{component analysis results}, detailing UI element types and their functionalities; and the \textbf{complete UI images}, providing visual context. Together, these components give the MLLM a holistic understanding of both the structure and visual design, enabling accurate and functionally complete UI code generation. \autoref{fig:synthesized_prompt} illustrates an example of the synthesized prompt.

The resulting prompt is fed into the MLLM for the initial code generation, setting the stage for the subsequent iterative code refinement process, where the generated code undergoes navigation and compilation checks for further improvement.

\subsection{Iterative Code Refinement}
\label{section:refinement}

The initial code generated by MLLMs may fail to implement the navigation logic defined in the PTG fully, contain syntax errors, or exhibit logical inconsistencies. To address these potential issues and ensure high-quality code output, \tool{} employs a two-pronged strategy for code refinement: \textbf{navigation consistency validation} and \textbf{compile-time error correction}. This iterative process systematically identifies and rectifies discrepancies between the generated code and the intended app structure and behavior.

\noindent\textbf{Navigation Consistency Validation.}
To ensure fidelity to the PTG, \tool{} implements a comprehensive navigation consistency validation process using static code analysis techniques. This process begins by parsing the generated code to extract all implemented navigation logic, specifically identifying framework-specific navigation statements without executing the code. The extracted navigation paths are then meticulously compared against the PTG definition, with a primary focus on identifying any missing transitions that are present in the PTG but absent in the generated code. This static analysis approach allows for additional navigation options in the code that may not be explicitly defined in the PTG, providing flexibility for potential enhancements or alternative user flows. Upon completion of the analysis, a detailed report is generated, highlighting any navigation inconsistencies, specifically focusing on missing transitions. This report serves as a crucial input for the MLLM in subsequent refinement iterations, offering targeted guidance to address specific navigation-related omissions and ensuring comprehensive implementation of the intended user journey as defined in the PTG.

\noindent\textbf{Compile-Time Error Correction.}
An integral part of \tool{}'s iterative refinement process is leveraging compiler feedback to detect and correct syntactic errors in the generated code. This is achieved through an automated compilation checking script that systematically processes the code using the appropriate development tools. The script captures and analyzes compiler error messages generated during this process. To bridge the gap between the compiler's technical output and the MLLM's natural language processing capabilities, these error messages are processed and restructured into clear, actionable instructions. This process involves extracting key information from error messages, including file names, line numbers, and error descriptions. The extracted information is then combined with relevant code snippets to form a structured prompt. This prompt is fed back to the MLLM, enabling it to make targeted corrections to the code.

\noindent\textbf{Iterative Refinement Process.} 
The code refinement process in \tool{} is iterative, designed to continuously enhance the quality of the generated code. Initially, \tool{} generates code for each app page based on the PTG. Then, it undergoes iterative cycles to check and correct navigation inconsistencies and compile-time errors. After generating each page's code, a navigation consistency check is performed, producing a report of any missing navigation paths. This report is fed back into the MLLM to address the gaps, with up to three iterations per page to ensure completeness. Once all pages are generated and compiled into a project, \tool{} conducts a compile-time error analysis. It identifies the relevant files and error messages, which are passed to the MLLM for targeted corrections. The MLLM refines the code based on both navigation and compile-time issues. This process is repeated, with compile-time error correction limited to three iterations for the entire project to maintain efficiency. This iterative approach ensures continuous improvement, addressing specific issues while keeping the process efficient through iteration limits.

\section{Evaluation}
To comprehensively evaluate the performance of \tool{}, we designed a series of experiments to address three key research questions (RQs):

\begin{description}
    \item[RQ1: Effectiveness.] How effective is \tool{}? How does \tool{} perform in comparison to baseline methods when operating autonomously without human intervention?
    
    \item[RQ2: Ablation Study.] How do individual key modules contribute to the overall performance of \tool{}?

    \item[RQ3: User Study and Manual Repair.] How do professional developers evaluate \tool{}'s generated UI code compared to the baseline in real-world scenarios? To what extent is manual effort required to repair compilation failures in \tool{}'s output?
\end{description}

\subsection{Dataset}
\label{section:dataset construct}
We curated a dataset of 50 mobile app UI designs (totaling 250 app pages), including 25 UI design mockups and 25 app screenshots, across 10 functional categories. This diversity allows us to assess the generalizability and robustness of our approach across various UI design paradigms and app types. We now provide details on the data sources, selection criteria, and dataset composition.

\noindent\textbf{Data Sources.}
Our dataset is drawn from two main sources: Figma~\cite{figma2024} and Google Play~\cite{GooglePlayStore2024}. Figma, a leading collaborative design tool, provides a selection of contemporary, high-quality mobile UI design mockups. Google Play, the official app distribution platform for Android devices, serves as our source for downloading apks, which we subsequently install on an emulator to manually capture screenshots. 
By incorporating cutting-edge designs from Figma and real-world apps from Google Play, we ensure that our approach is evaluated against a dataset that reflects both ideal design practices and practical implementation challenges in the mobile app development landscape.

\noindent\textbf{Selection Criteria.}
To ensure diversity in our dataset, we employed strict selection criteria. We classified mobile apps into 10 functional categories, including \textit{Social Networking}, \textit{Entertainment}, \textit{Productivity}, \textit{Education}, \textit{Wellness}, \textit{Finance}, \textit{Shopping \& E-commerce}, \textit{Travel \& Navigation}, \textit{News \& Information}, and \textit{Utility \& Practical Tools}. For Figma designs, we selected 194 mobile UI design mockups with over 300 followers, indicating their recognition within the design community. We then excluded designs that didn't meet our page count requirements (i.e., less than five pages) or weren't specifically designed for mobile apps. For Google Play apps, we focused on those ranking in the top 10 by download count within their respective categories, ensuring the representation of popular and widely used apps. We excluded paid apps and those requiring complex registration processes.
From each source, we selected 3-5 designs or apps per category that met our criteria. This range was established to accommodate potential scarcity in certain categories while maintaining a minimum threshold for diversity.
For each category, we ultimately collected no fewer than five designs; for those with more than five, we randomly selected five for experimentation to control experimental costs while ensuring research quality and statistical significance.

\noindent\textbf{Dataset Composition.}
The composition of our dataset is structured to provide a balanced and comprehensive representation of mobile UI designs. We selected 50 apps in total, equally divided between Figma designs and Google Play apps. These 50 apps are distributed across 10 functional categories, with five apps per category. The categorization was designed to encompass the broad spectrum of mobile app functionalities, aiming to provide a comprehensive representation of the app ecosystem.  For each app, we captured five logically connected pages, allowing us to analyze the flow and interaction between related interfaces.

\subsection{Experimental Setup}

\subsubsection{Evaluation Metrics}
\label{sec:metrics}

Our evaluation employs a set of metrics to assess the performance of \tool{}, encompassing visual fidelity, navigation logic, code quality, and practical usability.

To evaluate visual fidelity, we utilize two metrics:
\begin{itemize}
    \item \textbf{CLIP score}~\cite{si2024design2code}: This metric quantifies the semantic similarity between the generated and original UIs, providing an indicator of how well the generated UI captures the intended visual elements and overall design concept.
    \item \textbf{SSIM (Structural Similarity Index Measure) score}~\cite{ssimscore}: It assesses the layout and compositional accuracy, focusing on the spatial arrangement and structural similarities between the generated and original UIs.
\end{itemize}

The completeness of navigation logic implementation is measured by \textbf{PTG Coverage Rate (PCR)}. 
This metric, manually calculated by human evaluators, assesses the alignment between the implemented app structure and the intended design. It compares the predefined PTG (e.g., \autoref{fig:ptg_example}) with the UI Transition Graph (UTG) of the apps generated by \tool{}. The UTG is obtained using DroidBot~\cite{10.1109/ICSE-C.2017.8}, a widely adopted automated testing tool. The PCR is calculated as:
\[
PCR = \frac   {|\text{PTG}_{\text{edges}} \cap \text{UTG}_{\text{edges}}|}{|\text{PTG}_{\text{edges}}|} \times 100\%
\]

As described in \autoref{section:refinement}, we apply a compile-time error correction process, allowing up to three compilation attempts for each app. This iterative process aims to maximize the number of successfully compiled apps while minimizing manual intervention. We then employ different metrics based on the compilation outcome.

For successfully compiled apps (within three iterations), we calculate:
\begin{itemize}
    \item \textbf{Compilation Success Rate (CSR)}~\cite{wu2024uicoderfinetuninglargelanguage}: This measures the percentage of generated app code that is compiles successfully without errors after our automated iterative refinement process, without any human intervention: 
    \[
    CSR = \frac{\text{Number of successfully compiled apps}}{\text{Total number of apps}} \times 100\%
    \]
    \item \textbf{Average Compilation Iteration Count (ACIC)}: This metric indicates code generation efficiency and is computed as:
    \[
    ACIC = \frac{\sum_{i=1}^{n} iterations_i}{n}
    \]
    where $n$ is the total number of successfully compiled apps and $iterations_i$ is the number of compilation attempts for the $i$-th app (maximum 3). A lower ACIC suggests more efficient initial code generation.

\end{itemize}

For apps that fail to compile after three iterations (termed \textit{compilation-failed apps}), we measure:
\begin{itemize}
    \item \textbf{Average Manual Correction Time (AMCT)}: This quantifies the average time required for human developers to manually correct remaining errors in the generated code:
    \[
    AMCT = \frac{\text{Total manual correction Time}}{\text{Number of \textit{compilation-failed apps}}}
    \]
    AMCT provides insight into the practical efficiency gains of our method and the potential reduction in developer workload.
\end{itemize}

To provide a comprehensive assessment of the generated code's practical usability and efficiency, we employ \textbf{user study metrics}~\cite{Chen_2024}. These metrics evaluate:
\begin{itemize}
    \item \textbf{Code Availability}: Quantifies the proportion of usable code in the generated output.
    \item \textbf{Modification Time}: Measures time to adapt generated code to production standards.
    \item \textbf{Readability}: Assesses the structural understandability of the code.
    \item \textbf{Maintainability}: Measures the ease of adapting the code to new requirements.
\end{itemize}

Scores for these metrics are assigned using predefined scales and intervals, allowing for a nuanced evaluation of the code's quality from a developer's perspective.

\subsubsection{Baselines}
\label{sec:baselines}
We compare \tool{} against two baselines: GPT-4o and Claude-3.5 (specifically, Claude-3.5-Sonnet-20240620). Both baselines use unprocessed UI design images without our custom prompting techniques or additional processing.
To ensure a fair comparison and avoid underestimating the capabilities of MLLMs, we meticulously designed the prompts for the baselines. These prompts adhere to established practices from existing image-to-code research~\cite{wan2024automaticallygeneratinguicode,wu2024uicoderfinetuninglargelanguage,xiao2024prototype2codeendtoendfrontendcode} and widely-adopted techniques for LLM utilization~\cite{thedecoder2024,dong2024selfcollaborationcodegenerationchatgpt}. Both baseline approaches are evaluated within the React Native framework.

\subsubsection{Implementation Details}
Our experiments cover three popular mobile development frameworks that employ a declarative UI paradigm, i.e., React Native, Flutter, and ArkUI.
For each framework, we generate UI code using our \tool{} approach and evaluate its performance. All experiments were conducted on a server equipped with dual NVIDIA GeForce RTX 3090 GPUs, 64-core AMD EPYC processors, and 512GB of RAM. The code generation process was automated with custom Python scripts, totaling 2,935 lines, to minimize human intervention and bias.

\subsection{Effectiveness (RQ1)}
\label{sec:rq1}

To address RQ1, we conducted experiments on a carefully curated dataset comprising 50 app UI designs, each containing five pages, as stated in \autoref{section:dataset construct}. Given their status as state-of-the-art MLLMs, we selected GPT-4o and Claude-3.5 as base models to power \tool{}. The experiment's input consisted of these 50 app UI designs, while the output was the code generated by \tool{} for each app's five pages.
We integrated the generated code directly into an initial project in the IDE (i.e., Android Studio for React Native) without any manual intervention. The project was then packaged into an APK, which we installed and ran on the Android Studio built-in emulator to render the UI. We then evaluated the results according to the metrics outlined in \autoref{sec:metrics}.

As shown in \autoref{tab:comprehensive_metrics}, the evaluation results reveal that \tool{} (Claude-3.5) outperforms \tool{} (GPT-4o) in several key areas. For example, \tool{} (Claude-3.5) achieved a remarkable 98\% Compilation Success Rate (CSR), meaning that only one out of the 50 apps failed to compile automatically. In contrast, \tool{} (GPT-4o) had six apps that didn't compile under automated conditions. Upon analyzing the single app that failed to compile with \tool{} (Claude-3.5), we found that the issue was due to the app's use of non-existent file resources, which caused it to fail multiple compilation checks.

While \tool{} (GPT-4o) showed a marginally lower Average Compilation Iteration Count (ACIC) – 10 iterations for 44 compiled apps versus 11 for 49 apps with \tool{} (Claude-3.5) – this advantage was minimal, differing by only 0.01 iterations per app. Despite this slight edge in iteration efficiency, \tool{} (GPT-4o) lagged behind in other crucial metrics.  The SSIM score showed a particularly significant lead of 17.2\% (0.68 vs. 0.58), indicating that \tool{} (Claude-3.5) excels in accurately reproducing the structural elements of the original UI. Moreover, it exhibited a better PTG Coverage Rate (PCR), suggesting a more comprehensive understanding of navigation logic.
These results suggest that while both models show promise, \textbf{\tool{} (Claude-3.5) offers a more robust and accurate solution for automating UI code generation from designs}. Its higher CSR, coupled with better visual fidelity and navigation logic understanding, positions it as the more reliable choice for practical applications. The marginally higher ACIC for \tool{} (Claude-3.5) is a small trade-off for the significant improvements in other areas, particularly given the importance of visual accuracy and functional correctness in UI development.

\begin{table*}[h]
    \centering
    \caption{Comprehensive performance metrics across different methods and frameworks.}
    \resizebox{0.7\linewidth}{!}{%
    \begin{tabular}{lcccccc}
        \toprule[1.2pt]
        \textbf{Framework} & \textbf{Method} & \textbf{CLIP Score} & \textbf{SSIM Score} & \textbf{PCR (\%)} & \textbf{ACIC} & \textbf{CSR (\%)}  \\
        \hline
        \multirow{4}{*}{React Native} & Baseline (GPT-4o) & 0.66 & 0.44  & 1.4 & - & 76.0 \\
        & Baseline (Claude-3.5) & 0.65 & 0.41 & 43.3 & - & 74.0 \\
        & \tool{} (GPT-4o) & 0.87 & 0.58 & 88.6 & 0.23 & 88.0 \\
        & \tool{} (Claude-3.5) & 0.88 & 0.68 & 96.8 & 0.22 & 98.0 \\
        \hline
        Flutter& \tool{} (Claude-3.5) & 0.85  & 0.68 & 99.4 & 0.82 & 92.0 \\
        ArkUI& \tool{} (Claude-3.5) & 0.85 & 0.66 & 95.9 & 1.62 & 86.0 \\
        
    \bottomrule[1.2pt]
    \end{tabular}
    }
    \label{tab:comprehensive_metrics}
\end{table*}

\noindent\textbf{\tool{} vs. Baselines.}
As detailed in \autoref{sec:baselines}, we carefully selected and configured our baselines for comparison. The results, presented in \autoref{tab:comprehensive_metrics}, reveal significant limitations in using GPT-4o and Claude-3.5 directly for declarative UI code generation from mobile app UI designs.
Under fully automated conditions, the baseline methods achieved a CSR of 76\% (GPT-4o) and 74\% (Claude-3.5). Compilation failures primarily stemmed from issues such as using non-existent packages or components, inadequate special character handling, and component utilization without proper package imports.
While Baseline (GPT-4o) showed a slight edge in visual fidelity compared to Baseline (Claude-3.5), the advantage was marginal, with both falling short of our \tool{}'s performance. Notably, there was a stark contrast in PCR between the baselines, with Baseline (GPT-4o) achieving 1.4\% and Baseline (Claude-3.5) reaching 43.4\%. This disparity highlights Claude-3.5's substantial advantage in logical comprehension over GPT-4o in the baseline scenario.

Comparing \tool{} to the baselines reveals significant improvements across key metrics. The PCR, a crucial indicator of an app's navigational structure implementation, demonstrates \tool{}'s substantial advantage. In the React Native framework, \tool{} achieves 88.6\% and 96.8\% coverage with GPT-4o and Claude-3.5 respectively, far surpassing the corresponding baseline methods (1.4\% and 43.3\%). This marked improvement suggests \tool{}'s enhanced ability to capture and implement complex app structures accurately.
The compilation statistics further underscore \tool{}'s superiority in code quality. The 98\% CSR of \tool{} (Claude-3.5) compared to Baseline (Claude-3.5)'s 74\%, and \tool{} (GPT-4o)'s 88\% versus Baseline (GPT-4o)'s 76\%, clearly demonstrate \tool{}'s significant outperformance of baseline methods. This high CSR, coupled with low iteration requirements (only 11 iterations needed for 49 successfully compiled apps), indicates that \tool{} generates more robust, deployment-ready code. 
Based on the comprehensive performance of Claude-3.5 and GPT-4o in both baseline scenarios and with \tool{}, we have selected Claude-3.5 as the base model for our subsequent experiments.

\noindent\textbf{Generalization Capabilities.} 
The results for Flutter and ArkUI presented in \autoref{tab:comprehensive_metrics} demonstrate \tool{}'s robust performance across diverse UI frameworks, highlighting its impressive generalization capabilities and adaptability.
Consistent with our evaluation approach for React Native, we compiled Flutter projects in Android Studio and ArkUI projects in DevEco Studio, integrating the generated code directly into initial projects without manual intervention. 

Flutter emerges as particularly noteworthy, with \tool{} achieving an exceptional PCR of 99.4\%. This near-perfect score indicates \tool{}'s profound understanding and accurate implementation of navigation logic within the Flutter ecosystem. React Native and ArkUI also display impressive results, with PCRs of 96.8\% and 95.9\% respectively, further reinforcing \tool{}'s versatility across different framework paradigms.
The CSRs across frameworks, ranging from 86\% to 98\%, reveal both \tool{}'s strengths and areas for potential improvement. 
Specifically, the slightly lower CSR for ArkUI presents an intriguing avenue for future investigation. We explore potential strategies for improving \tool{}'s performance with ArkUI in \autoref{sec:RAG}, where we discuss the implementation of advanced techniques to potentially bridge this performance gap.

\find{\textbf{Answer to RQ1:}
\tool{} demonstrates high effectiveness in generating functional UI code across multiple frameworks, significantly outperforming baselines. It achieves up to 98\% CSRs, superior visual fidelity, and navigation logic implementation (PCR up to 99.4\%). With strong generalization across React Native, Flutter, and ArkUI, and Claude-3.5 as the preferred base model, \tool{} proves to be a robust solution for automated UI development, despite some room for improvement in ArkUI, which will be discussed in \autoref{sec:RAG}.}

\subsection{Ablation Study (RQ2)}
\label{sec:rq2}

To assess the contribution of each key component in \tool{}, we conducted two ablation studies on a subset of 13 apps from our dataset. This selection was based on CLIP score distributions across app categories, ensuring diverse representation. Our sample includes: three apps from categories with the lowest median CLIP scores, five from categories with median performance, three from categories showing high CLIP score variability, one from a category with low variability, and one from a category exhibiting multiple outliers in CLIP scores. This diverse sample spans various complexities and visual effects, providing a robust basis for evaluating \tool{}'s performance across different scenarios. We focused on two key aspects: the impact of CV techniques, and the role of PTG and the refinement process. It's important to note that we couldn't isolate PTG construction alone due to its inherent coupling with the iterative refinement process, as the navigation checks during refinement rely on the PTG. These studies help isolate the effects of these components on the overall performance of \tool{}.

\begin{table*}[h]
    \centering
    \caption{Comprehensive performance metrics across different \tool{} versions.}
    \resizebox{0.7\linewidth}{!}{%
    \begin{tabular}{lccccc}
    \toprule[1.2pt]
        \textbf{Version} & \textbf{CLIP Score} & \textbf{SSIM Score} & \textbf{PCR (\%)} & \textbf{ACIC} & \textbf{CSR (\%)} \\
        \hline
        \tool{} (Full) & 0.90 & 0.72 & 96.8 & 0.22 & 98.0 \\
        \tool{} (without CV) & 0.88 & 0.57 & 95.1 & 0 & 100.0 \\
        \tool{} (without PTG+Refinement) & 0.87 & 0.68 & 23.1 & - & 80.0 \\
    \bottomrule[1.2pt]
    \end{tabular}
    }
    \label{tab:ablation_comprehensive_metrics}
\end{table*}

\noindent\textbf{Impact of CV techniques on UI Generation.}
In this ablation study, we removed the UI component extraction and representation module while retaining the PTG and refinement processes to investigate the impact of CV techniques on UI generation. The results in \autoref{tab:ablation_comprehensive_metrics} illustrate that the absence of the UI component extraction and representation module led to a significant decrease in visual similarity scores, with the SSIM score dropping from 0.72 to 0.57, and the CLIP score slightly decreasing from 0.90 to 0.88. These reductions in structural and semantic similarity metrics directly reflect CV techniques' impact on the MLLM's component recognition capabilities. The decrease in visual similarity scores without the combination of CV techniques indicates that MLLMs struggle more with accurately identifying and reproducing individual UI elements, as evidenced by the drop in PCR from 96.8\% to 95.1\%.

\noindent\textbf{Impact of PTG and Refinement Process.}
In the second ablation study, we removed both the PTG and refinement components, relying solely on the UI component extraction and representation for UI generation. Although the visual similarity scores remained relatively high based on \autoref{tab:ablation_comprehensive_metrics}, the PCR dropped drastically to 23.1\% and the CSR decreased to 80.0\%. These results emphasize the critical role of PTG and refinement in ensuring accurate navigation logic, code compilability, and overall UI quality.

\find{\textbf{Answer to RQ2:} The combination of CV techniques contributes to the visual accuracy of UI generation, while the PTG and refinement processes ensure functional correctness and high code quality. The synergistic integration of these three components forms the core technology of \tool{}, enabling the generation of high-fidelity and functionally-correct UIs.}

\subsection{User Study and Manual Repair (RQ3)}
\label{sec:rq3}

\begin{figure}[htbp]
    \centering
    \subfigure[Code Availability]{
        \centering
        \includegraphics[width=0.48\textwidth]{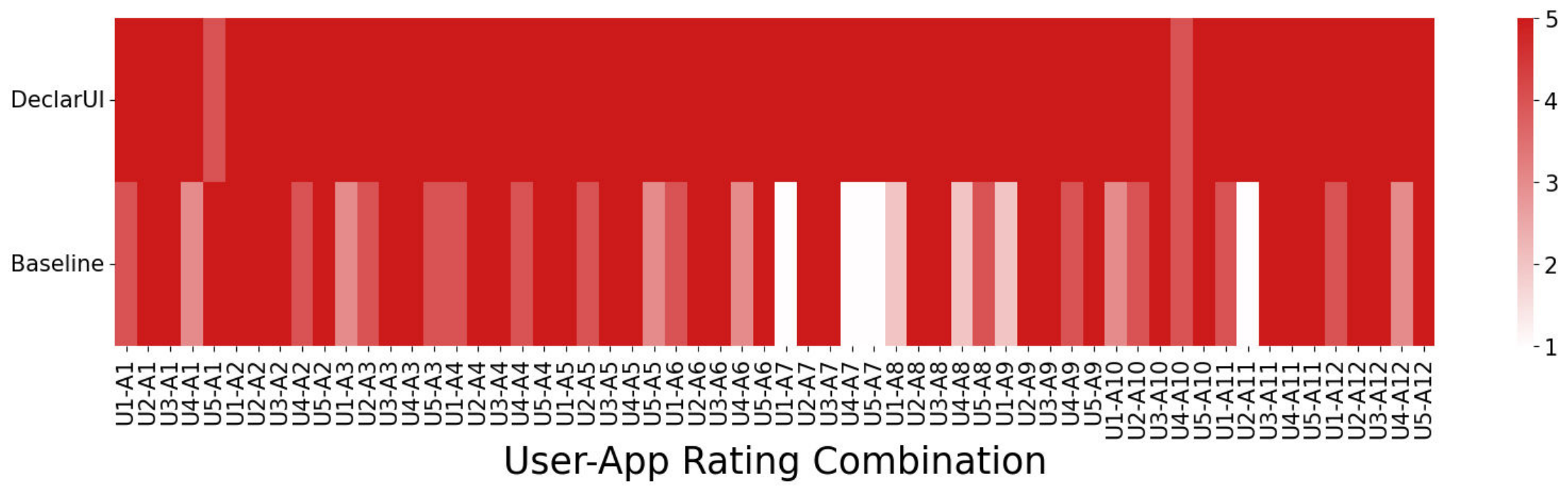}
        \label{fig:availability}}
    \subfigure[Modification Time]{
        \centering
        \includegraphics[width=0.48\textwidth]{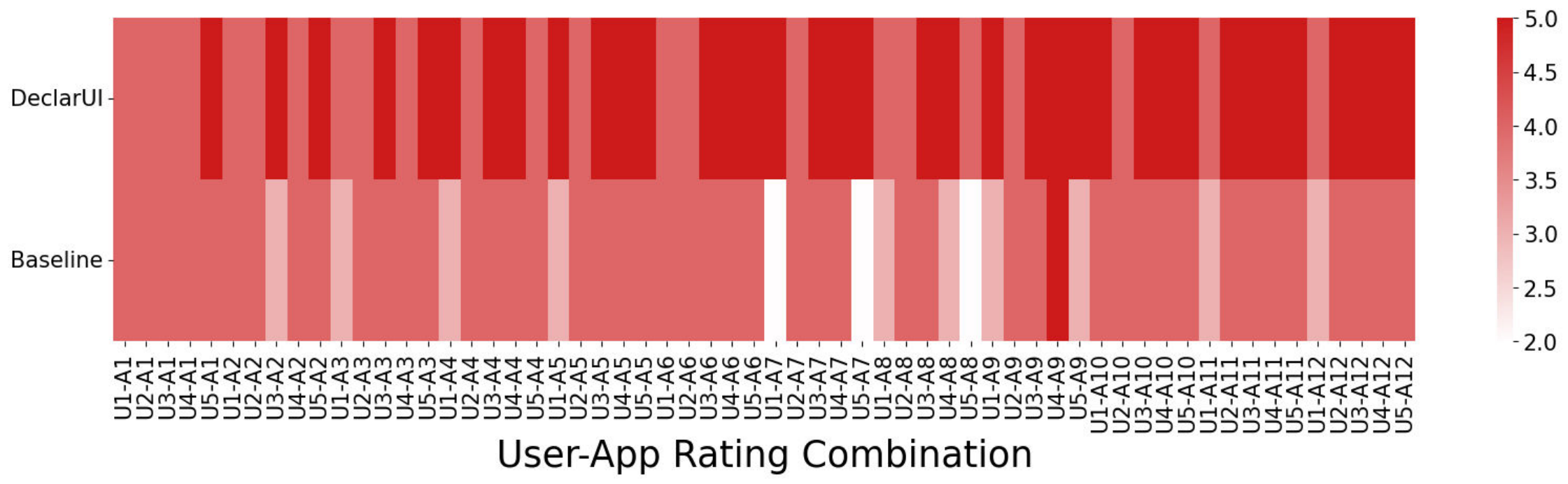}
       
        \label{fig:modification}}

    \subfigure[Readability]{
        \centering
        \includegraphics[width=0.48\textwidth]{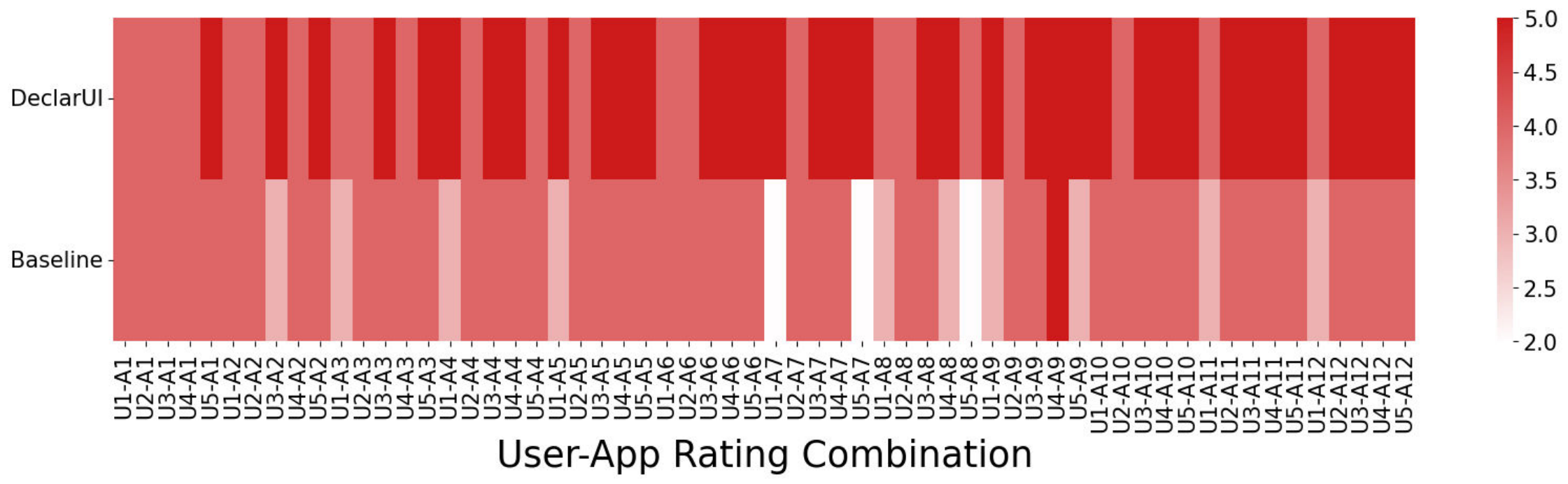}
      
        \label{fig:readaility}}
    \subfigure[Maintainability]{
        \centering
        \includegraphics[width=0.48\textwidth]{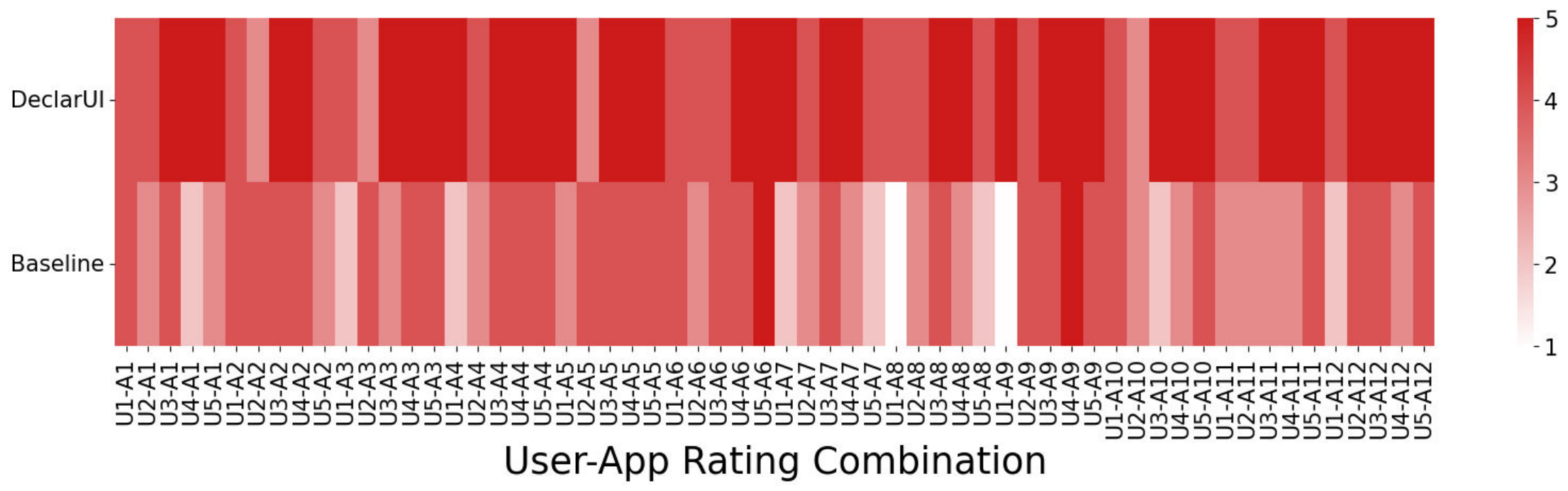}
        \label{fig:maintainability}}
    
    \caption{Heatmaps of user study metrics.}
    \label{fig:heatmap}
\end{figure}

\subsubsection{User Study}
\label{sec:user study}
To comprehensively evaluate the performance of our automatic generation method, we focused not only on the visual quality of the rendered UI but also on the overall quality of the generated code. In real-world application scenarios, particularly in industry settings, code readability, availability, and maintainability are three crucial evaluation dimensions~\cite{chen2022uilayersmergermerging,Chen_2024}. To assess these aspects, we conducted a user study with experienced developers.

\noindent\textbf{Procedures.}
We recruited five React Native developers with an average of over two years of experience. All participants were compensated \$50 for their time and expertise. Each participant reviewed 12 apps with UI code generated by \tool{} and Baseline (Claude-3.5). These apps were carefully selected based on their CLIP scores: three well-performing apps (CLIP score > 0.85), six average-performing apps (0.8 < CLIP score < 0.85), and three poor-performing apps (CLIP score < 0.8), all randomly chosen within their respective groups. 
For code availability evaluation, participants modified each set of UI page code to meet basic business requirements. We tracked modifications using git, following the method of Chen et al.~\cite{Chen_2024}, mapping scores from 0.8 with each 5\% interval corresponding to levels 1-5. After modifying each set, participants rated its readability and maintainability on a five-point Likert scale. Code modification time measures the time required to adjust the code to production standards. We scored this on 5-minute intervals, with 5 points for completion within 5 minutes and 1 point for over 20 minutes. To avoid tool bias, participants used Android Studio, with a 20-minute time limit per app. We recorded modification time and collected ratings for each set of UI page code. This study design allowed us to evaluate the generated code's quality through both objective measures (modification time and extent) and subjective assessments (readability and maintainability ratings) from experienced professionals.

\begin{table*}[h]
\centering
\caption{Participant ratings of \tool{} vs. Baseline across different metrics.}
\resizebox{0.7\linewidth}{!}{
\begin{tabular}{@{}cccccc@{}}
\toprule[1.2pt]
\textbf{Metric}              & \textbf{Code Availability} & \textbf{Modification Time} & \textbf{Readability} & \textbf{Maintainability} \\ \midrule
Baseline Score      & 4.15              & 2.45              & 3.75        & 3.37            \\
\tool{} Score       & 4.97              & 4.03              & 4.62        & 4.55            \\
P-value             & 9.23E-08          & 2.20E-10          & 2.57E-13    & 1.33E-12        \\ 
\bottomrule[1.2pt]
\end{tabular}}
\label{tab:human evaluation}
\end{table*}

\noindent\textbf{Results.}
We present the average scores and p-values using the Mann-Whitney U test in \autoref{tab:human evaluation}, while more detailed statistical analyses are visualized using a set of heatmaps in \autoref{fig:heatmap}. Professional React Native engineers consistently rated \tool{}-generated code significantly higher in terms of availability, modifiability, readability, and maintainability. The substantial improvements observed in all these metrics were statistically significant (p < 0.05), underscoring the robustness of our findings. Notably, \tool{} achieved a near-perfect score in code availability (4.97 out of 5), demonstrating its exceptional performance in generating immediately usable code. The marked reduction in modification time (4.03 for \tool{} vs. 2.45 for the baseline) indicates that \tool{}-generated code requires considerably less effort to adapt to production standards. This is further evidenced by the fact that users were able to complete modifications on 73.3\% of \tool{}-generated code within 10 minutes. In contrast, 80\% of the baseline code required more than 10 minutes to modify, and the actual modification time for the baseline is likely underestimated, as \textbf{27.1\% of this 80\% was not completed even within 20 minutes}. Furthermore, the superior readability (4.62 vs. 3.75) and maintainability (4.55 vs. 3.37) scores highlight the enhanced long-term value of the code produced by \tool{}.

\begin{figure}[htbp]
\centering
\subfigure[Original UI design]{
\label{fig:original_ui}
\includegraphics[width=0.25\linewidth]{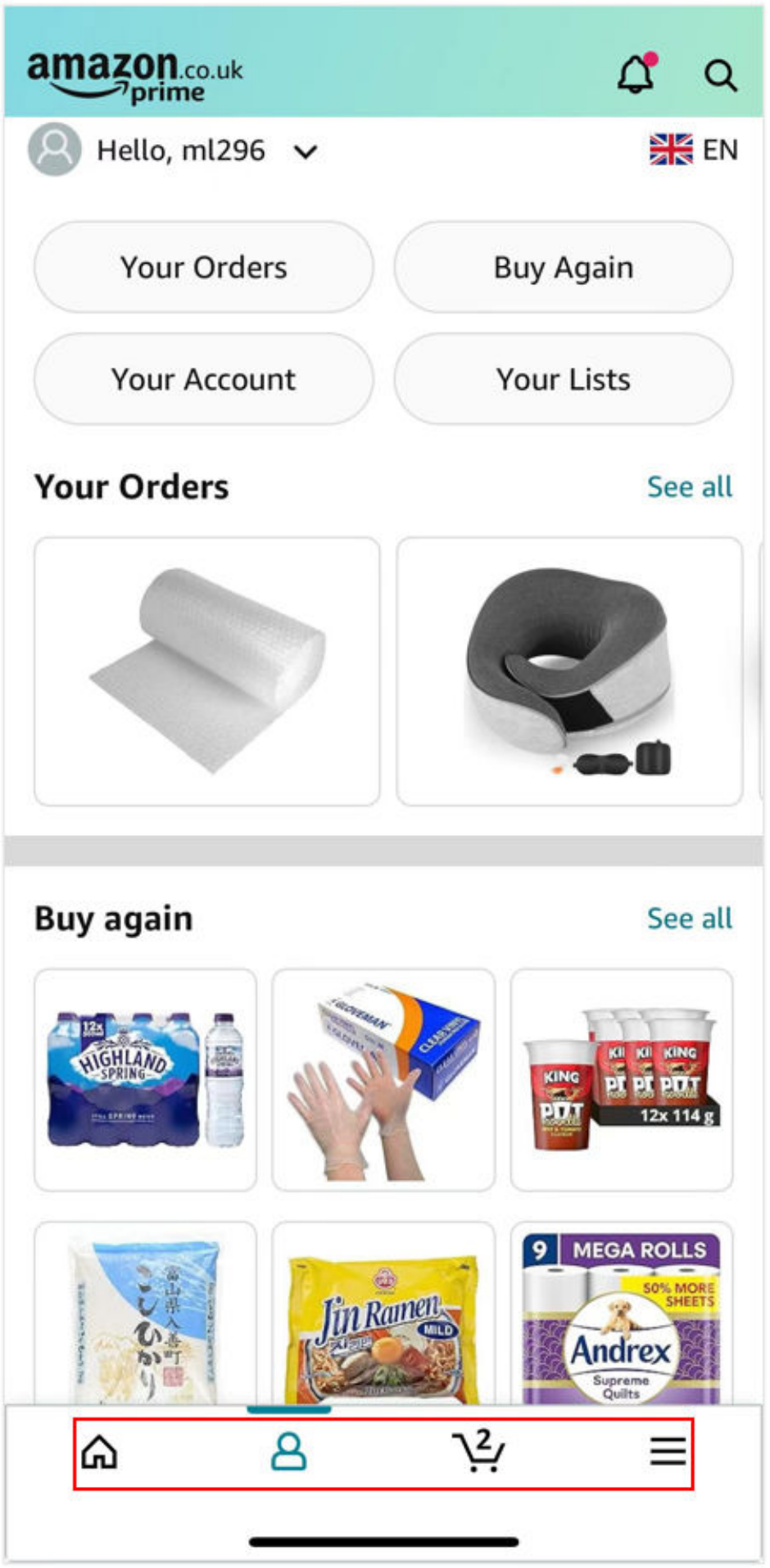}}
\hspace{0.1cm}
\subfigure[Baseline]{
\label{fig:baseline_ui}
\includegraphics[width=0.25\linewidth]{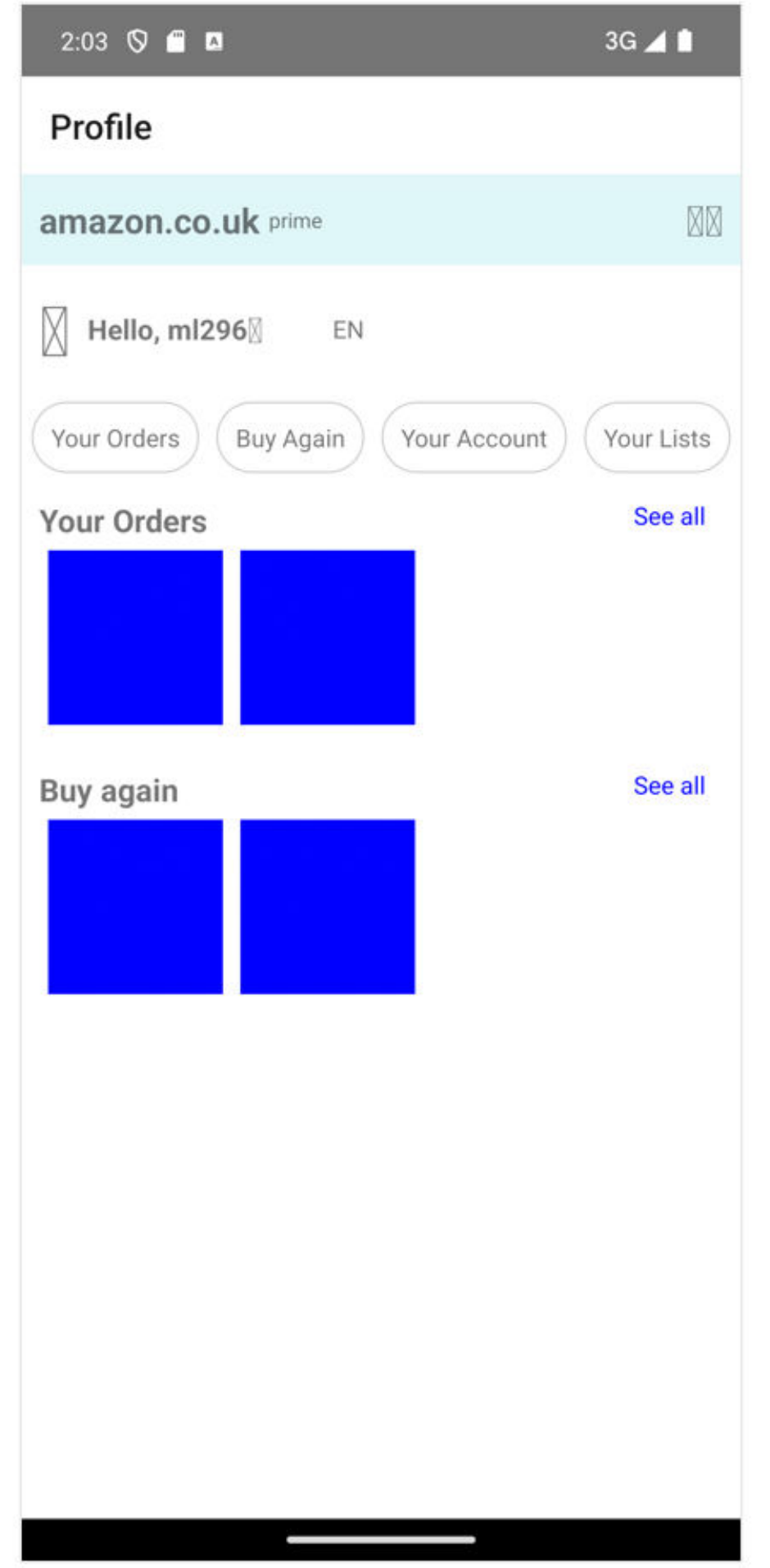}}
\hspace{0.1cm}
\subfigure[\tool{}]{
\label{fig:declarui_ui}
\includegraphics[width=0.25\linewidth]{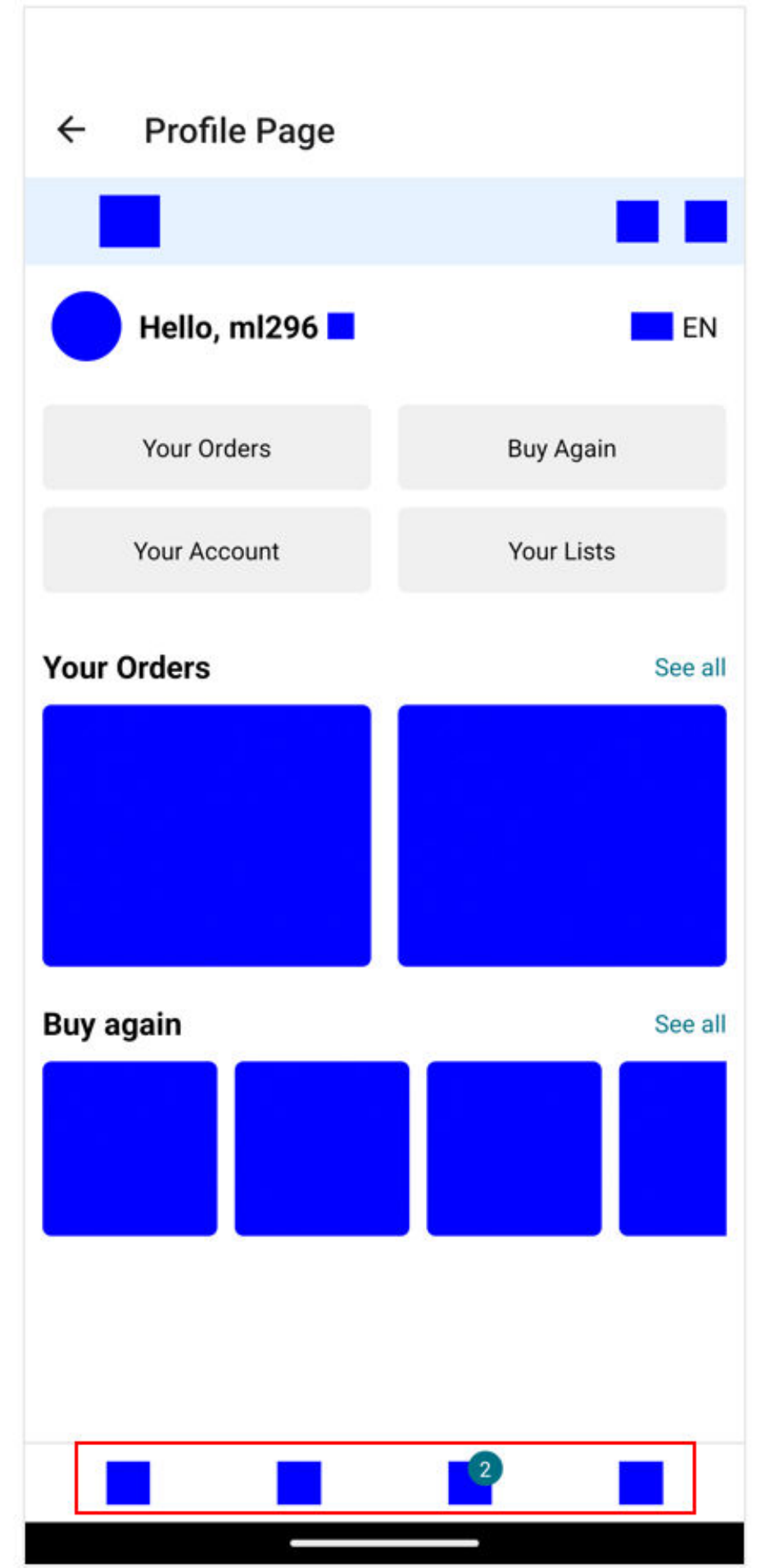}}
\caption{The original UI design image, output from Baseline (Claude-3.5), and \tool{}'s generated result.}
\label{fig:component_issue}
\end{figure}

\begin{figure}[htbp]
\centering
\subfigure[\tool{} UTG]{
\label{fig:declarui_utg}
\includegraphics[width=0.8\linewidth]{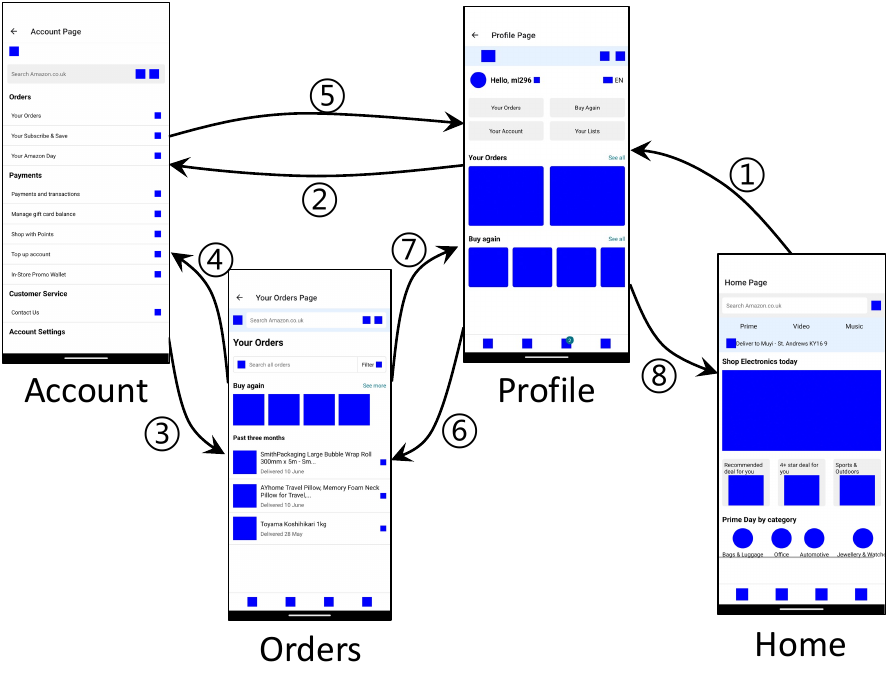}}
\hspace{0.15cm}
\subfigure[Baseline (Claude-3.5) UTG]{
\label{fig:baseline_utg}
\includegraphics[width=0.7\linewidth]{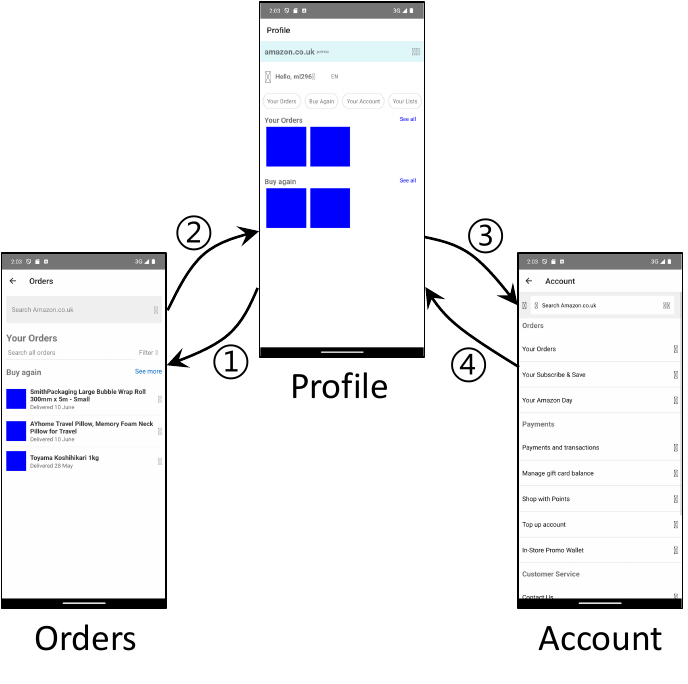}}
\caption{Navigation completeness \& interaction consistency.}
\label{fig:navigation_issue}
\end{figure}

\noindent\textbf{Case Study.}
To demonstrate \tool{}'s superiority, we present concrete examples in \autoref{fig:component_issue} and \autoref{fig:navigation_issue}. Our analysis reveals three primary areas where \tool{} excels over baseline methods: component fidelity, navigation completeness, and interaction consistency. We illustrate these using the \texttt{com.amazon.mShop.android. shopping} app~\cite{AmazonApp2024}. \tool{} achieves superior \textbf{component fidelity} by accurately replicating all elements from the original design (\autoref{fig:declarui_ui}), including the bottom navigation bar omitted by the baseline (\autoref{fig:baseline_ui}). This completeness extends to \textbf{navigation completeness} (\autoref{fig:navigation_issue}). \tool{} maintains comprehensive navigation pathways (\autoref{fig:declarui_utg}), addressing gaps present in the baseline's structure (\autoref{fig:baseline_utg}). Consequently, our approach demonstrates improved \textbf{interaction consistency}, particularly in navigation between related UI components. This consistency is evident in the uniform navigation patterns to ``Orders'' from both ``Profile'' and ``Account'' pages (\autoref{fig:declarui_utg}), ensuring a more coherent user experience.

\subsubsection{Manual Repair}

To comprehensively assess the human effort required to repair compilation failures of \tool{}, we used the Average Manual Correction Time (AMCT) metric. Our analysis revealed that for the React Native framework, \tool{} (Claude-3.5) required an AMCT of 48s, while \tool{} (GPT-4o) needed 43.2s. In comparison, Flutter required 62s, and ArkUI took 183s. These results demonstrate \tool{}'s efficiency in generating code that requires minimal manual intervention, highlighting its superiority across different frameworks.

\find{\textbf{Answer to RQ3:} \tool{} significantly outperforms the baseline in code quality, usability, and efficiency. The user study and case study demonstrate \tool{}'s superiority in generating high-quality, readily usable UI code that closely matches original designs. Notably, \tool{} requires minimal manual intervention, a characteristic observed consistently across multiple UI frameworks.}

\section{Discussion}
\subsection{Retrieval-Augmented Generation (RAG): Impact and Implications}
\label{sec:RAG}

The performance of \tool{} across the three frameworks is uneven, with ArkUI showing relatively poor performance compared to the others (\autoref{tab:comprehensive_metrics}). This suboptimal performance likely results from the MLLM's limited knowledge of ArkUI, a newer and less mainstream framework, compared to its more comprehensive understanding of established frameworks.

To address this challenge, we explored integrating a knowledge graph (KG)~\cite{peng2023knowledgegraphsopportunitieschallenges} to supplement the MLLM's understanding of ArkUI. This approach follows a Retrieval-Augmented Generation (RAG)~\cite{fan2024surveyragmeetingllms} method to compensate for the MLLM's lack of knowledge about less mainstream frameworks. When generating UI code, we first query the KG and then incorporate the retrieved information into the prompt. We tested this approach with 20 apps with UI designs from our dataset, totaling 100 app pages. The results showed promising improvements: ACIC is relatively reduced by 0.52 (1.1 vs. 1.62), CSR increased by 4\% (90\% vs. 86\%), and AMCT reduced by 139.5s (43.5s vs. 183s). While our preliminary results demonstrated potential, the improvements were not as significant as anticipated, primarily due to time constraints limiting the sophistication of our KG implementation.

Future research should focus on refining the RAG-based approach, particularly for less common frameworks like ArkUI. This could involve developing a more comprehensive and sophisticated KG that captures the intricacies of ArkUI syntax and design patterns. Additionally, exploring techniques to dynamically update the KG with new framework features and best practices could ensure the system remains current and effective. Further investigation into optimizing the integration of KG-retrieved information with the MLLM's existing knowledge could potentially yield more substantial performance improvements. Moreover, extending this approach to other emerging UI frameworks could enhance the versatility and adaptability of \tool{}.

\subsection{Threat to Validity}
\label{sec:thrests}

\textbf{Internal Validity.} The primary threats stem from our evaluation metrics and method implementation across different frameworks. CLIP Score and SSIM, used for UI similarity assessment, may not fully capture human-perceived UI similarity, potentially overlooking subtle differences crucial to user experience. PTG Coverage, employed for navigation logic accuracy, ensures implementation of expected page transitions but may not reflect the actual user experience of navigation fluidity and intuitiveness. Fortunately, our user study confirmed the validity of our evaluations, demonstrating that \tool{} indeed outperformed the baselines.

\noindent\textbf{External Validity.} Our study faces several threats related to the performance of MLLM influenced by prompt design and the experience levels of user study participants. Different prompts could lead to varying results, making it challenging to isolate the true impact of our approach. Although we selected participants with similar backgrounds, inherent biases may still affect the results. Fortunately, \autoref{fig:heatmap} showed that participants' opinions on the same app were relatively consistent.
Another threat involves the scope and size of our UI design sample. Despite efforts to include diverse UI designs within the mobile app domain, our sample may not fully represent all mobile app UI types. The limited sample size of 50 UI design sets, due to MLLMs' token cost constraints, may not capture the full variability of mobile app UIs. Future research should expand the size and diversity of the UI sample to ensure comprehensive coverage of mobile app UI types.

\section{Related Work}

\noindent\textbf{Mobile App UI-to-Code Generation.}
Mobile app UI-to-code translation has seen various approaches in recent years. \cite{beltramelli2017pix2codegeneratingcodegraphical,encoder-decoder} were limited by traditional UI frameworks and lacked support for multi-page apps. \cite{wan2024automaticallygeneratinguicode} explored code generation from natural language descriptions but struggled with complex designs and fine-grained component recognition. \cite{GUI_skeleton,object_detect,screenparsing,iconlabel} introduced intermediate representations to aid UI code generation, yet still required significant manual effort to produce the final code.
Unlike previous works, \tool{} is the first to focus on serving declarative UI frameworks, combining advanced CV techniques for precise component recognition with PTG-based inter-page logic capture to enable fully automated, multi-page mobile app UI code generation.

\noindent\textbf{Image-to-Code Generation.}
Image-to-code generation research, particularly in UI generation, has advanced significantly through three main approaches: Deep Learning-based methods~\cite{beltramelli2017pix2codegeneratingcodegraphical,Cai2023ANC,10.1007/s10462-023-10631-z}, CV-based methods~\cite{jain2019sketch2codetransformationsketchesui,screenshot_to_code, Baul2020RecentPI}, and Multimodal Learning methods~\cite{liu2024luminamgptilluminateflexiblephotorealistic,Kwak2023AMD,Luo2021ASO}.
\cite{beltramelli2017pix2codegeneratingcodegraphical} pioneered deep learning techniques to generate code from GUI screenshots. CV-based methods, exemplified by \cite{jain2019sketch2codetransformationsketchesui}, converted hand-drawn sketches into HTML code, while \cite{screenshot_to_code} advanced this by leveraging GPT-4V and DALL·E~3 to transform screenshots into functional code.
Addressing limitations of single-modal approaches, Multimodal Learning methods emerged. \cite{liu2024luminamgptilluminateflexiblephotorealistic} presented a multimodal autoregressive model for various vision and language tasks, including code generation from images. \cite{Kwak2023AMD} demonstrated how combining different data types can enhance model robustness and versatility, potentially improving accuracy and contextual understanding in code generation tasks.

\section{Conclusion}

In this paper, we presented \tool{}, an approach that combines CV techniques, MLLMs, and iterative compiler-driven optimization to generate high-quality declarative UI code from design mockups and screenshots. Our evaluation demonstrates that \tool{} outperforms state-of-the-art methods across multiple metrics, including PTG coverage rates, visual similarity, average compilation iteration counts, and compilation success rates. \tool{}'s multi-framework compatibility is evidenced by its ability to generalize across React Native, Flutter, and ArkUI. The positive feedback from professional developers in our user study underscores \tool{}'s practical value. By addressing challenges in component recognition, interactive logic understanding, and code reliability, \tool{} represents a significant advancement in bridging UI design and implementation.

\balance

\bibliographystyle{ACM-Reference-Format}
\bibliography{main}

\end{document}